\documentclass[journal]{IEEEtran}
\usepackage{cite}
\usepackage{amsmath,amssymb,amsfonts}
\usepackage{algorithmic}
\usepackage{graphicx}
\usepackage{textcomp}
\usepackage{xcolor}

\usepackage[utf8]{inputenc} 
\usepackage[T1]{fontenc}    
\usepackage{hyperref}       
\usepackage{url}            
\usepackage{booktabs}       
\usepackage{amsfonts}       
\usepackage{nicefrac}       
\usepackage{microtype}      

\usepackage{bm}
\usepackage{comment}
\usepackage{amsmath}
\usepackage{enumerate}
\usepackage{enumitem}
\usepackage{mathrsfs}
\usepackage{makeidx}         
\usepackage{graphicx}        
\usepackage{multicol}        
\usepackage{subfigure}
\usepackage{epsfig,amssymb,latexsym}
\usepackage{psfrag}
\usepackage{algorithmic}
\usepackage{algorithm}

\usepackage{fancyhdr}
\usepackage{multirow}

\usepackage{color}

\newcommand{\mb}{\mathbb}
\newcommand{\mr}{\mathrm}
\newcommand{\mc}{\mathcal}

\ifCLASSINFOpdf
\else
\fi

\begin{document}
%


\title{Data Poisoning Attacks on Federated Machine Learning}

\author{Gan~Sun, 
        Yang~Cong,~\IEEEmembership{Senior Member,~IEEE,}
        Jiahua~Dong,
        Qiang~Wang
        and~Ji~Liu  
\IEEEcompsocitemizethanks{\IEEEcompsocthanksitem G. Sun, Y. Cong, J. Dong and Q. Wang are with State Key Laboratory of Robotics, Shenyang Institute of Automation, Chinese Academy of Sciences, Shenyang, 110016, China, Email: sungan1412@gmail.com, congyang81@gmail.com, dongjiahua@sia.cn, wangqiang@sia.cn. \protect\\
\vspace{-8pt}
\IEEEcompsocthanksitem J. Dong is also with University of Chinese Academy of Sciences, Beijing. \protect
\IEEEcompsocthanksitem J. Liu is with the Department of Computer Science, University of Rochester, Rochester, NY 14627 USA (e-mail: jliu@cs.rochester.edu).
}
\thanks{Manuscript received April 19, 2005; revised August 26, 2015.}}
%
%

\markboth{Journal of \LaTeX\ Class Files,~Vol.~14, No.~8, August~2015}%
{Shell \MakeLowercase{\textit{et al.}}: Bare Demo of IEEEtran.cls for Computer Society Journals}
%

\IEEEtitleabstractindextext{%
\begin{abstract}
Federated machine learning which enables resource-constrained node devices (\emph{e.g.,} mobile phones and IoT devices) to learn a shared model while keeping the training data local, can provide privacy, security and economic benefits by designing an effective communication protocol. However, the communication protocol amongst different nodes could be exploited by attackers to launch data poisoning attacks, which has been demonstrated as a big threat to most machine learning models. In this paper, we attempt to explore the vulnerability of federated machine learning. More specifically, we focus on attacking a federated multi-task learning framework, which is a federated learning framework via adopting a general multi-task learning framework to handle statistical challenges. We formulate the problem of computing optimal poisoning attacks on federated multi-task learning as a bilevel program that is adaptive to arbitrary choice of \emph{target} nodes and \emph{source attacking} nodes. Then we propose a novel systems-aware optimization method, \underline{ATT}ack on \underline{F}ederated \underline{L}earning (AT$^2$FL), which is efficiency to derive the implicit gradients for poisoned data, and further compute optimal attack strategies in the federated machine learning. Our work is an earlier study that considers issues of data poisoning attack for federated learning. To the end, experimental results on real-world datasets show that federated multi-task learning model is very sensitive to poisoning attacks, when the attackers either directly poison the \emph{target} nodes or indirectly poison the related nodes by exploiting the communication protocol.


\end{abstract}

}

\maketitle

\IEEEdisplaynontitleabstractindextext

\IEEEpeerreviewmaketitle

\section{Introduction}\label{sec:introduction}
\IEEEPARstart{M}achine learning has been widely-applied into a broad array of applications, \emph{e.g.,} spam filtering~\cite{zhao2016optimizing} and natural gas price prediction~\cite{alfeld2016data}. Among these applications, the reliability or security of the machine learning system has been a great concern, including adversaries~\cite{huang2011adversarial,yuan2019adversarial}. For example, for product recommendation system~\cite{wang2017coupled}, researchers can either rely on public crowd-sourcing platform, \emph{e.g.,} Amazon Mechanical Turk or Taobao, or private teams to collect training datasets. However, both of these above methods have the opportunity of being injected corrupted or poisoned data by attackers. To improve the robustness of real-world machine learning systems, it is critical to study how well machine learning performs under the poisoning attacks.

For the attack strategy on machine learning methods, it can be divided into two categories: causative attacks and exploratory attacks~\cite{barreno2010security}, where exploratory attacks influence learning via controlling over training data, and exploratory attacks can take use of misclassifications without affecting training. However, previous researches on poisoning attacks focus on the scenarios that training samples are collected in a centralized location, or the training samples are sent to a centralized location via a distributed network, \emph{e.g.,} support vector machines~\cite{biggio2012poisoning}, autoregressive models~\cite{alfeld2016data} and collaborative filtering~\cite{li2016data}. However, none of the current works study poisoning attacks on federated machine learning\cite{yang2019federated,xu2019verifynet,konevcny2015federated}, where the training data are distributed across multiple devices (\emph{e.g.,} users' mobile devices: phones/tablets), and may be privacy sensitive. To further improve its robustness, in this paper, our work explores how to attack the federated machine learning.

\begin{figure}[t]
 \hspace{-2.0mm}
   \subfigure{
       \label{fig:lifelong} 
       \centering
        \begin{minipage}{0.48\textwidth}
         \centering
    \includegraphics[width=250pt, height=140pt]{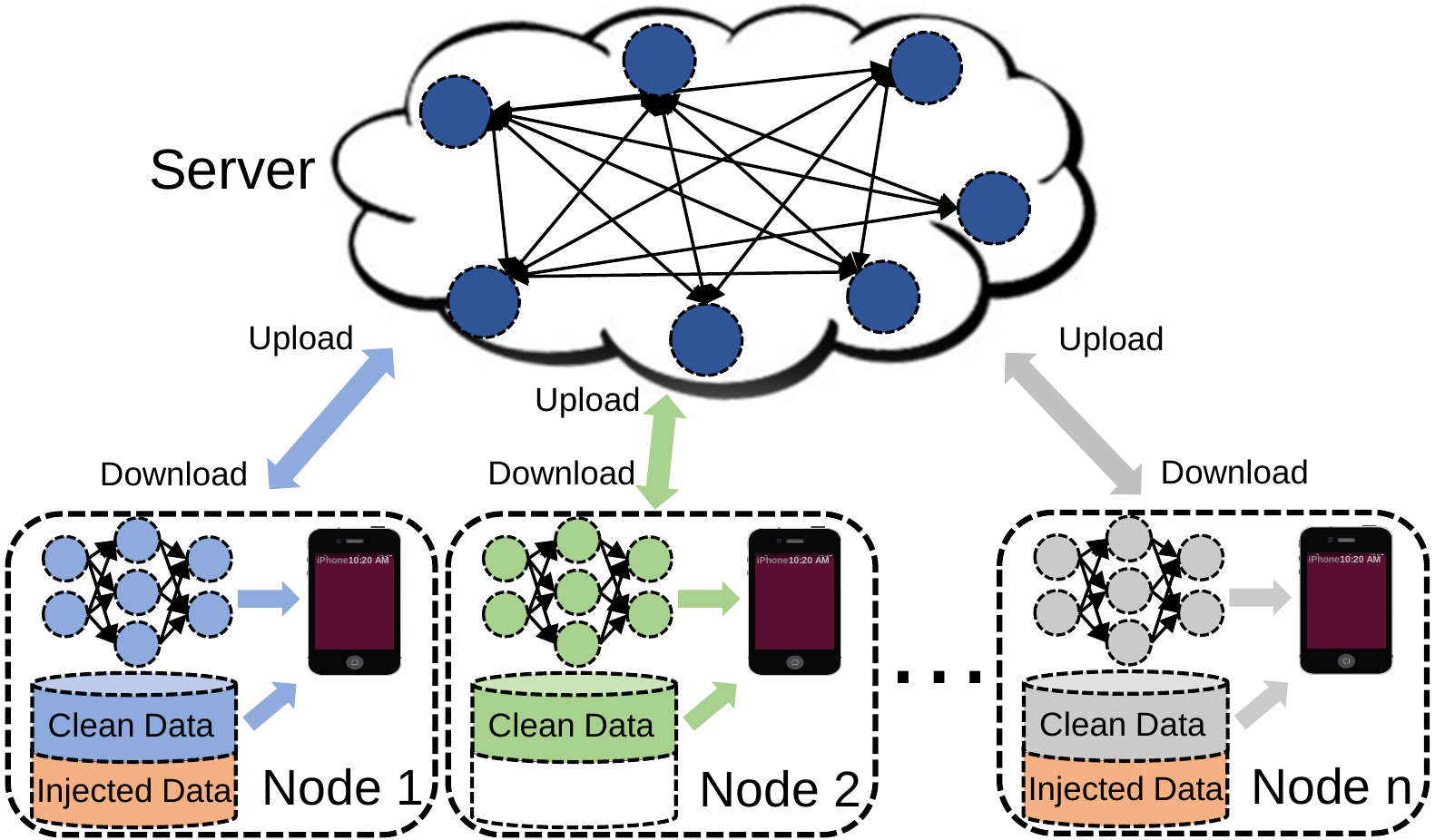}
         \end{minipage}}
   \caption{The demonstration of our data poisoning attack model on federated machine learning, where different colors denote different nodes, and there are $n$ nodes in this federated learning system. Some nodes are injected by corrupted/poisoned data, and some nodes are only with clean data.}
  \label{fig:DataPoisoningFederatedLearning}
 \end{figure}


For the federated machine learning, its main idea is to build machine learning models based on data sets that are distributed across multiple devices while preventing data leakage. Although recent improvements have been focusing on overcoming the statistical challenges (\emph{i.e.,} the data collected across the network are in a \emph{non-IID} manner, where the data on each node are generated by a distinct distribution) or improving privacy-preserving, the attempt that makes federated learning more reliability under poisoning attacks, is still scarce. For example, consider several different e-commerce companies in a same region, and the target is to establish a prediction model for product purchase based on user and product information, \emph{e.g.,} user's browsing and purchasing history. The attackers can control a  prescribed number of user accounts and inject the poisoning data in a direct manner. Furthermore, due to the communication protocol existing amongst different companies, this protocol also opens a door for the attacker to indirectly affect the unaccessible \emph{target} nodes, which is also not addressed by existing poisoning methods whose training data are collected in a centralized location.


Motivated by the aforementioned analysis, we attempt to analyze optimal poisoning attacks on federated machine learning. More specifically, as shown in Figure~\ref{fig:DataPoisoningFederatedLearning}, we focus on the recently-proposed federated multi-task learning framework~\cite{smith2017federated}, a federated learning framework which captures node relationships among multiple nodes to tackle the statistical challenges in the federated setting. Our work is to formulate the optimal poisoning attack strategy on federated multi-task learning model as a general bilevel optimization problem, which can be adaptive to any choice of \emph{target} nodes and \emph{source attacking} nodes. However, current optimization methods for this bilevel problem are not suited to tackle the systems challenges (\emph{e.g.,} high communication cost, stragglers) that exist in federated learning. As a key component of this work, we thus design a novel optimization method, \underline{ATT}ack on \underline{F}ederated \underline{L}earning (AT$^2$FL), to derive the implicit gradients for computing poisoned data in the \emph{source attacking} nodes. Furthermore, the obtained gradient can be effectively used to compute the optimal attack strategies.
Finally, we empirically evaluate the effectiveness of the proposed optimal attack strategy against random baselines on
several real-world datasets. The experiment results strongly support our proposed model when attacking federated machine learning based on the communication protocol.

\indent The novelty of our proposed model is threefold:
\begin{itemize}
  \item We propose a bilevel optimization framework to compute optimal poisoning attacks on federated machine learning. To our best knowledge, this is an earlier attempt to explore the vulnerability of federated machine learning from the perspective of data poisoning.
  \item We derive an effective optimization method, \emph{i.e.,} \underline{ATT}ack on \underline{F}ederated \underline{L}earning (AT$^2$FL), to solve the optimal attack problem, which can address systems challenges associated with federated machine learning.
  \item We demonstrate the empirical performance of our optimal attack strategy, and our proposed AT$^2$FL algorithm with several real-world datasets. The experiment results indicate that the communication protocol among multiple nodes opens a door for attacker to attack federated machine learning.
\end{itemize}


\section{Related Work}\label{sec:related work}
Our work mainly draws from data poisoning attacks and federated machine learning, so we give a brief review on these two topics.

For the \textbf{data poisoning attacks}, it has become an urgent research field in the adversarial machine learning, in which the target is against machine learning algorithms~\cite{barreno2006can,huang2011adversarial}. The earlier attempt that investigates the poisoning attacks on support vector machines (SVM)~\cite{biggio2012poisoning}, where the adopted attack uses a gradient ascent strategy in which the gradient is obtained based on properties of the SVM's optimal solution. Furthermore, poisoning attack is investigated on many machine learning models, including autoregressive models~\cite{alfeld2016data}, matrix factorization based collaborative filtering~\cite{li2016data} and neural networks for graph data~\cite{zugner2018adversarial}. In addition to single task learning models, perhaps \cite{zhao2018data} is the most relevant work to ours in the context of data poisoning attacks, which provides the first study on one much challenging problem, \emph{i.e.,} the vulnerability of multi-task learning. However, the motivations for \cite{zhao2018data} and our work are significantly different as follows:
\begin{itemize}
  \item The data sample in \cite{zhao2018data} are put together, which is different from the scenario in federated machine learning, \emph{i.e.,} machine learning models are built based on datasets that are distributed across multiple nodes/devices while preventing data leakage.
  \item The proposed algorithm in \cite{zhao2018data} is based on optimization method of current multi-task learning methods, which is not suited to handle the systems challenges in federated learning, including high communication cost, etc. Handling these challenges in the setting of data poisoning attacks is a key component of this work.
\end{itemize}


\indent For the \textbf{federated machine learning}, its main purpose is to update classifier fast for modern massive datasets, and the training data it can handle are with the following properties~\cite{konevcny2016federated}: 1) \emph{Non-IID}: data on each node/device may be drawn from a different distribution; 2) \emph{Unbalanced}: the number of training samples for different nodes/devices may vary by orders of magnitude. Based on the distribution characteristics of the data, federated learning~\cite{yang2019federated} can be categorized into: 1) horizontal (sample-based) federated learning, \emph{i.e.,} datasets share the same feature space but different in samples. The representative work is a multi-task style federated learning system~\cite{smith2017federated}, which is proposed to allow multiple nodes to complete separate tasks while preserving security and sharing knowledge; 2) vertical (feature-based) federated learning, \emph{i.e.,} two datasets share the same sample ID space but differ in feature space. Several privacy-preserving machine learning methods have been presented for vertically partitioned data, \emph{e.g.,} secure linear regression~\cite{gascon2016secure}, gradient descent methods~\cite{han2009privacy}; 3) federated transfer learning, \emph{i.e.,} two datasets differ not only in samples but also in feature space. For this case, traditional transfer learning techniques can be adopted into provide solutions for the entire sample and feature space with the federated setting. As an first attempt, \cite{bagdasaryan2018backdoor} develops a new model-replacement methodolody that exploits these vulnerabilities and demonstrated its efficacy on federated learning tasks. However, its objective is to retain high accuracy on the backdoor subtask after attacking. On the contrast, in this work, we try to fill data poisoning gap by investigating poisoning attack against horizontal federated machine learning.

\section{The Proposed Formulation}\label{sec:formulation}
In this section, we introduce the details of our proposed poisoning attacks strategies on federated machine learning. Therefore, we firstly present some preliminaries about federated multi-task learning, which is a general multi-task learning formulation in federated setting. Then we provide the mathematical form of our poisoning attack formulation, followed by how to optimize our proposed model.

\begin{table}[t]
\caption{Notations for all the used Variables.}
\centering
\scalebox{1.05}{
\begin{tabular}{|c|c|}
\hline
 {Variables}&Interpretation \\
 \hline\hline
${D}_{\ell}$  & Clean data for the $\ell$-th node \\ \hline
${D}_{t}$  & Clean data for the $t$-th \emph{target} node \\ \hline
$\hat{D}_{\ell}$  & Injected data for the $\ell$-th node \\ \hline
$\hat{D}_{s}$  & Injected data for the $s$-th \emph{source attacking} node \\ \hline
$\mc{N}_{\mr{tar}}$ & A set of \emph{target} nodes    \\ \hline
$\mc{N}_{\mr{sou}}$ & A set of \emph{source attacking} nodes    \\ \hline
$\mc{H}$ & Upper level function in Eq.~\eqref{eq:ours}  \\ \hline
$W$ & Weight matrix    \\ \hline
$\Omega$  & Model relationship among nodes  \\ \hline

 \end{tabular}
}
\label{table:result}
\end{table}

\subsection{Federated Machine Learning}
In the setting of federated machine learning, the target is to learn a model over data that resides on, and has been generated by, $m$ distributed nodes. Each node/device, $\ell\in [m]$, may generate local data via a distinct distribution, and so it is natural to fit separate models, $\{w_1,\ldots,w_m\}$, to the distributed data-one for each local dataset. In this paper, we focus on an effective horizontal (sample-based) federated learning model, \emph{i.e.,} federated multi-task learning\cite{smith2017federated}, which is proposed via incorporating with a general multi-task learning framework. More specifically, the federated multi-task learning model can be formulated as:
\begin{equation}\label{eq:fmtl}
  \min_{W,\Omega}\; \sum_{\ell=1}^m\sum_{i=1}^{n_{\ell}} \mc{L}_{\ell}(w_{\ell}^{\top}x_{\ell}^i,y_{\ell}^i)+\lambda_1\mr{tr}(W\Omega W^{\top})+\lambda_2\left\|W\right\|_F^2,
\end{equation}
where $(x_{\ell}^i,y_{\ell}^i)$ is the $i$-th sample for the $\ell$-th node, $n_{\ell}$ is the number of clean samples for the $\ell$-th node, $W=[w_1,\ldots,w_m]\in\mb{R}^{d\times m}$ is a matrix whose $\ell$-th column is the weight vector for the $\ell$-th node and matrix $\Omega\in \mb{R}^{m\times m}$ models relationships among nodes. $\lambda_1>0$ and $\lambda_2>0$ are the parameters to control regularization terms.

From the equation above, we can notice that solving for matrix $\Omega$ is not dependent on the data and therefore can be computed centrally; one major contribution of \cite{smith2017federated} is an efficient distributed optimization method for the $W$ step. Furthermore, the update of $W$ can be achieved by extending works on distributed primal-dual optimization~\cite{shalev2014accelerated}. Let $n:=\sum_{\ell=1}^m n_{\ell}$ and $X:=\mr{Diag} (X_1,\ldots,X_m) \in\mb{R}^{md\times n}$. With the fixed variable $\Omega$, the dual formulation of optimization problem in Eq.~\eqref{eq:fmtl}, defined with respect to dual variables $\alpha_{\ell}\in \mb{R}^{n_{\ell}}$, is given by:
\begin{equation}\label{eq:dual_alpha}
  \min_{\alpha,W,\Omega}\; \sum_{\ell=1}^m\sum_{i=1}^{n_{\ell}} \mc{L}_{\ell}^*(-\alpha_{\ell}^i)+\lambda_1\mc{R}^*(X\alpha),
\end{equation}
where $\mc{L}_{\ell}^*$ and $\mc{R}^*$ are the conjugate dual functions of $\mc{L}_{\ell}$ and $\mr{tr}(W\Omega W^{\top})+\lambda_2/\lambda_1\left\|W\right\|_F^2$, respectively, and $\alpha_{\ell}^i$ in $\alpha\in\mb{R}^n$ is the dual variable of the $i$-th data point $(x_{\ell}^i,y_{\ell}^i)$ for the $\ell$-th node. Meanwhile, we denote ${D}_{\ell}=\{({X}_{\ell},{\alpha}_{\ell},{y}_{\ell})|{X}_{\ell}\in\mb{R}^{d\times {n}_{\ell}},{\alpha}_{\ell}\in\mb{R}^{{n}_{\ell}},{y}_{\ell}\in\mb{R}^{{n}_{\ell}}\}$ as the clean data in the node $\ell$ in this work.


\subsection{Poisoning Attacks on Federated Multi-task Learning}
In this section, we firstly introduce the problem setting of the data poisoning attack on federated machine learning. Then three kinds of attacks based on real-world scenarios are provided, followed by a bilevel formulation to compute optimal attacks.

Suppose that the attacker aims to degrade the performance of a set of \emph{target} nodes $\mc{N}_{\mr{tar}}\subset m$ by injecting corrupted/poisoned data to a set of \emph{source attacking} nodes $\mc{N}_{\mr{sou}}\subset m$. Based on the dual problem in Eq.~\eqref{eq:dual_alpha}, we denote  $\hat{D}_{\ell}=\{(\hat{X}_{\ell},\hat{\alpha}_{\ell},\hat{y}_{\ell})|\hat{X}_{\ell}\in\mb{R}^{d\times \hat{n}_{\ell}},\hat{\alpha}_{\ell}\in\mb{R}^{\hat{n}_{\ell}},\hat{y}_{\ell}\in\mb{R}^{\hat{n}_{\ell}}\}$ as the set of malicious data injected to the node $\ell$, where $\hat{n}_{\ell}$ denotes the number of injected samples for the node $\ell$. More specifically, $\hat{D}_{\ell}$ will be $\emptyset$, \emph{i.e.,} $\hat{n}_{\ell}=0$, if ${\ell}\notin \mc{N}_{\mr{sou}}$. We also define the following three kinds of attacks based on real-world federated learning scenarios.
\begin{itemize}
  \item \textbf{Direct attack:} $\mc{N}_{\mr{tar}}=\mc{N}_{\mr{sou}}$. Attacker can directly inject data to all the \emph{target} nodes since a door will be opened when collecting data. For example, consider learning or recognizing the activities of mobile phone users in a cell network based on their individual sensor, text or image data. Attackers can directly attack the target mobile phones (nodes) by providing counterfeit sensor data into the target phones (nodes).
  \item \textbf{Indirect attack:} $\mc{N}_{\mr{tar}}\cap \mc{N}_{\mr{sou}}=\emptyset$. Attacker cannot directly inject data to any of the target nodes. However, due to the communication protocol existing amongst multiple mobile phones, the attacker can inject poisoned data to other mobile phones and affect the target nodes in an indirect way.
  \item \textbf{Hybrid attack:} An attack style which combines direct attack and indirect attack, \emph{i.e.,} the attacker can inject poisoned data samples into target nodes and source attacking nodes simultaneously.
\end{itemize}

To maximally degrade the performance of \emph{target} nodes, we then formulate the optimal attack problem as the following bilevel optimization problem by following \cite{biggio2012poisoning}:
\begin{equation}\label{eq:ours}
\begin{aligned}
 &  \max_{\{\hat{D}_{s}|s\in \mc{N}_{\mr{sou}}\}} \; \sum_{\{t|t\in \mc{N}_{\mr{tar}}\}} \mc{L}_{t}(D_t,w_t), \\
 & \quad\;\;  {s.t.,} \;  \min_{\alpha,W,\Omega}\sum_{\ell=1}^{m}\frac{1}{n_{\ell}+\hat{n}_{\ell}} \mc{L}_{\ell}^*(D_{\ell}\cap \hat{D}_{\ell}) + \lambda_1\mc{R}^*(X\alpha).
  \end{aligned}
\end{equation}
where $\hat{D}_{s}$ denotes the injected data for the $s$-th \emph{source attacking} node. Intuitively, the variables in the upper level problem are the data points $\hat{D}_{s}$, and we denote the upper level problem as a $\mc{H}$ in this paper. The lower level problem in the Eq.~\eqref{eq:ours} is a federated multi-task learning problem with training set consisting of both clean and injected data points. Therefore, the lower level problem can be considered as the constraint of the upper level problem.

\section{\underline{ATT}ack on \underline{F}ederated \underline{L}earning (AT$^2$FL)}
This section proposes an effective algorithm for computing optimal attack strategies, \emph{i.e.,} AT$^2$FL. More specifically, we follow the setting of most data poisoning attack strategies (\emph{e.g.,} \cite{biggio2012poisoning,zhao2018data}), and design a projected stochastic gradient ascent based algorithm that efficiently maximally increases the empirical loss of target nodes, and further damages their classification/regression performance. Notice that there is no closed-form relation between the empirical loss and the injected data, so we intend to compute the gradients by exploiting the optimality conditions of the subproblem.

\subsection{Attacking Alternating Minimization}
Bilevel problems are usually hard to optimize due to their non-convexity property. In our bilevel formulation Eq.~\eqref{eq:ours}, although the upper level problem is a simple primal problem, the lower level problem is highly non-linear and non-convex. To effectively solve this problem, the idea is to iteratively update the injected data in the direction of maximizing the function $\mc{H}$ of target nodes. Therefore, in order to reduce the complexity of the optimal attack problem, we need to optimize over the features of injected data $(\hat{x}_s^i,\hat{\alpha}_s^i)$ by fixing the labels of injected data. Then the update rules for injected data $\hat{x}_s^i$ can be written as:
\begin{equation}\label{eq:update_x}
\begin{aligned}
  (\hat{x}_s^i)^k  \leftarrow \mr{Proj}_{\mb{X}}\big((\hat{x}_s^i)^{k-1}+\eta_1\nabla_{(\hat{x}_s^i)^{k-1}}\mc{H}\big),
  \end{aligned}
\end{equation}
where $\eta_1>0$ is the step size, $k$ denotes the $k$-th iteration, and $\mb{X}$ represents the feasible region of the injected data, which is specified by the first constraint in the upper level problem $\mc{H}$. More specifically, $\mr{Proj}_{\mb{X}}(x)$ is $x$ if $\left\|x\right\|_2\leq r$; $xr/\left\|x\right\|_2$, otherwise, \emph{i.e.,} $\mb{X}$ can be considered as an $\ell_2$-norm ball by following~\cite{daubechies2008accelerated}. Accordingly, the corresponding dual variable $\hat{\alpha}_s^i$ can be updated gradually as the $\hat{x}_s^i$ comes as following:
\begin{equation}\label{eq:update_alphaWithx}
 (\hat{\alpha}_s^i)^k \leftarrow (\hat{\alpha}_s^i)^{k-1}+\Delta{(\hat{\alpha}_s^i)}.
\end{equation}

 \subsection{Gradients Computation}
To compute the gradient of the $t$-th target node in Eq.~\eqref{eq:update_x}, \emph{i.e.,} $\nabla_{\hat{x}_s^i}\mc{L}_t(D_t,w_t)$, we adopt the chain rule and then obtain the following equation:
\begin{equation}\label{eq:update_x_chainrule}
  \nabla_{\hat{x}_s^i}\mc{L}_t(D_t,w_t)=\nabla_{w_t}\mc{L}_t(D_t,w_t)\cdot \nabla_{\hat{x}_s^i} w_t.
\end{equation}
From the equation above, we can notice that the first term on the right side can be easily computed since it depends only on the loss function $\mc{L}_{t}(\cdot)$. However, the second term on the right side depends on the optimality conditions of lower level problem of Eq.~\eqref{eq:ours}. In the following, we show how to compute the gradient $ \nabla_{\hat{x}_s^i} w_t$ with respect to two commonly-adopted loss functions: least-square loss (regression problems) and hinge loss (classification problems). Please check the definitions in the \textbf{Appendix A}.

\renewcommand{\algorithmicrequire}{\textbf{Input:}}
\renewcommand{\algorithmicensure}{\textbf{Output:}}
\begin{algorithm}[t]
\caption{\underline{ATT}ack on \underline{F}ederated \underline{L}earning (AT$^2$FL)}
\begin{algorithmic}[1]
\REQUIRE  Nodes $\mc{N}_{\mr{tar}}, \mc{N}_{\mr{sou}}$, attacker budget $\hat{n}_s$;
\STATE  Randomly initialize $\forall \ell\in \mc{N}_{\mr{sou}}, $
  \begin{equation}\label{eq:initializing_attacker}
     \hspace{-20pt} \hat{D}_{\ell}^0=\{(\hat{X}_{\ell}^0,\hat{\alpha}_{\ell}^0,\hat{y}_{\ell}^0)|\hat{X}_{\ell}^0\in \mb{R}^{d\times \hat{n}_{\ell}},\hat{\alpha}_{\ell}^0\in\mb{R}^{\hat{n}_{\ell}},\hat{y}_{\ell}^0\in\mb{R}^{\hat{n}_{\ell}}\}
  \end{equation}
\STATE Initialize $\hat{D}_{\ell}=\hat{D}_{\ell}^0, \;\forall \ell\in \mc{N}_{\mr{sou}}$ and matrix $\Omega^0$;
\FOR  {$k=0,1,\ldots$}
\STATE      Set subproblem learning rate $\eta$;
      \FOR  {all nodes $\ell=1,2, \ldots, m $ in parallel}
 \STATE  Compute the approximate solution $\Delta \alpha_{\ell}$ via Eq.~\eqref{eq:closed-form_lsl} or Eq.~\eqref{eq:closed-form_hl};
 \STATE  Update local variables $\alpha_{\ell}\leftarrow \alpha_{\ell}+\Delta \alpha_{\ell}$;
     \IF    {$\ell\in \mc{N}_{\mr{sou}}$}
         \STATE  Compute the approximate solution $\Delta \hat{\alpha}_{\ell}$ via Eq.~\eqref{eq:closed-form_lsl} or Eq.~\eqref{eq:closed-form_hl};
         \STATE  Update local variables $\hat{\alpha}_{\ell}\leftarrow \hat{\alpha}_{\ell}+\Delta \hat{\alpha}_{\ell}$;
         \ENDIF
      \ENDFOR
      \STATE Update $W^k$ and $\Omega^k$ based on the latest $\alpha$;
\FOR {all source nodes $s=1,2,\ldots, \mc{N}_{\mr{sou}}$ in parallel}
\STATE Update $\hat{x}_s^i$ based on the Eq.~\eqref{eq:update_x};
\ENDFOR
\STATE $\hat{D}_{\ell}=\hat{D}_{\ell}^k, \;\forall \ell\in \mc{N}_{\mr{sou}}$;
\ENDFOR
\end{algorithmic}
\end{algorithm}



Based on the loss functions above, we first fix variable $\Omega$ to eliminate the constraints of the lower level problem, and  obtain the following dual subproblem:
\begin{equation}\label{eq:dual_w1}
\min_{\alpha,W}\; \sum_{\ell=1}^m\frac{1}{n_{\ell}+\hat{n}_{\ell}} \mc{L}_{\ell}^*(D_{\ell}\cup\hat{D}_{\ell})+\lambda_1\mc{R}^*(X\alpha).
\end{equation}
Since the optimality of the lower level problem can be considered as a constraint to the upper level problem, we can treat  $\Omega$ in problem of Eq.~\eqref{eq:dual_w1} as a constant-value matrix when computing the gradients. Additionally, since $\mc{R}^*(X\alpha)$ is continuous differentiable, $W$ and $\alpha$ could be connected via defining the next formulation by following~\cite{shalev2014accelerated}:
\begin{equation}\label{eq:define_w}
  w_{\ell}(\alpha)=\nabla \mc{R}^*(X\alpha).
\end{equation}
Therefore, the key component on the rest is how to update the dual variable $\alpha$ and further compute the corresponding gradients in Eq.~\eqref{eq:update_x_chainrule}.

\textbf{Update Dual Variable $\alpha$:} Consider the maximal increase of the dual objective with a least-square loss function or a hinge loss function, where we only allow to change each element of $\alpha$. We can reformulate Eq.~\eqref{eq:dual_w1} as the following constrained optimization problem for the $\ell$-th node:
\begin{equation}\label{eq:dual_w2}
  \min_{\alpha} \sum_{\ell=1}^m \frac{1}{n_{\ell}+\hat{n}_{\ell}} \big(\mc{L}_{\ell}^*(-\alpha_{\ell}^i)+\mc{L}_{\ell}^*(-\hat{\alpha}_{\ell}^{i'})\big)+\lambda_1\mc{R}^*(X[\alpha_{\ell};\hat{\alpha}_{\ell}]).
\end{equation}
To solve Eq.~\eqref{eq:dual_w2} across distributed nodes, we can define the following data-local subproblems, which are formulated via a careful quadratic approximation of this dual problem to separate computation across the nodes. Moreover, at every step $k$, two samples (\emph{i.e.,} $i$ in $\{1,\ldots,n_{\ell}\}$ and $i'$ in $\{1,\ldots,\hat{n}_{\ell}\}$) are chosen uniformly at random from original clean data and injected data, respectively, and the updates of both $\alpha_{\ell}^i$ and $\hat{\alpha}_{\ell}^{i'}$ in node $\ell$ can be computed as:
\begin{equation}\label{eq:update_alpha}
\begin{aligned}
(\alpha_{\ell}^i)^k &=(\alpha_{\ell}^i)^{k-1}+\Delta \alpha_{\ell}^i, \\
(\hat{\alpha}_{\ell}^{i'})^k &=(\hat{\alpha}_{\ell}^{i'})^{k-1}+\Delta \hat{\alpha}_{\ell}^{i'}, \\
\end{aligned}
\end{equation}
where both $\Delta \alpha_{\ell}^i$ and $\Delta \hat{\alpha}_{\ell}^{i'}$ are the stepsizes chosen to achieve maximal ascent of the dual objective in Eq.~\eqref{eq:dual_w2} when all variables are fixed. To achieve maximal dual ascent, one has to optimise:
\begin{equation}\label{eq:update_DeltaAlpha}
\begin{aligned}
\Delta \alpha_{\ell}^i & \!=\!\arg\min_{a\in \mb{R}} \mc{L}_{\ell}^*(-(\alpha_{\ell}^i+a))\!+\!a\langle w_{\ell}(\alpha_{\ell}), x_{\ell}^i\rangle\!+\!\frac{\lambda_1}{2}\!\left\|x_{\ell}^i a \right\|_{\rm{M_\ell}}^2  \\
\Delta \hat{\alpha}_{\ell}^{i'}&\!\!=\!\arg\min_{\hat{a}\in \mb{R}} \mc{L}_{\ell}^*(-(\hat{\alpha}_{\ell}^{i'}+\hat{a}))\!+\!\hat{a}\langle w_{\ell}(\hat{\alpha}_{\ell}), x_{\ell}^{i'}\rangle\!+\!\frac{\lambda_1}{2}\!\left\|x_{\ell}^{i'}\hat{a} \right\|_{\rm{M_\ell}}^2
\end{aligned}
\end{equation}
where $M_{\ell}\in\mb{R}^{d\times d}$ is the $\ell$-th diagonal block of a symmetric positive definite matrix $M$. Furthermore, $M^{-1}= \bar{\Omega}+\lambda_2/\lambda_1 {I}_{md\times md}$, where $\bar{\Omega}:=\Omega \otimes \mr{I}_{d\times d} \in \mb{R}^{md\times md}$. Therefore, $\Delta \alpha_{\ell}^i$ can be computed in a closed-form for the least-square loss function, \emph{i.e.,}
\begin{equation}\label{eq:closed-form_lsl}
  \Delta \alpha_{\ell}^i=\frac{y_{\ell}^i-(x_{\ell}^i)^{\top}x_{\ell}^i\alpha_{{\ell}}^i-0.5(\alpha_{\ell}^i)^{k-1}}{0.5+\lambda_1\left\|x_{\ell}^i\right\|_{\rm{M_{\ell}}}^2}.
\end{equation}
Meanwhile, $\Delta \hat{\alpha}_{\ell}^{i'}$ can be computed in a same manner. Furthermore, we substitute the hinge loss into the optimization problem in Eq.~\eqref{eq:dual_w1}, and can obtain:
\begin{equation}\label{eq:closed-form_hl}
\begin{aligned}
  \Delta \alpha_{\ell}^i=&y_{\ell}^i \mr{max} \Big( 0, \mr{min}\big(1, \!\frac{1-(x_{\ell}^i)^{\top}x_{\ell}^i(\alpha_{\ell}^i)^{k-1}y_{\ell}^i}{\lambda_1\left\|x_{\ell}^i\right\|_{M_{\ell}}^2}\!+\!(\alpha_{\ell}^i)^{k-1}y_{\ell}^i\big)\Big) \\ &-(\alpha_{\ell}^i).
  \end{aligned}
\end{equation}

\textbf{Update Gradient:} Given Eq.~\eqref{eq:closed-form_lsl}, Eq.~\eqref{eq:closed-form_hl} with Eq.~\eqref{eq:define_w}, we can compute the gradient in Eq.~\eqref{eq:update_x}. For the least-square loss, we can the gradient of each injected data $\hat{x}_s^i$ with its associated \emph{target} node $t$:
\begin{equation}\label{eq:update_x_lsl}
\begin{aligned}
  &  \nabla_{(\hat{x}_s^i)}\mc{L}_t((w_t)^{\top}x_t^j,y_t^j) \\
  =&2((w_t)^{\top}x_t^j-y_t^j)x_t^j \cdot \nabla_{\hat{x}_s^i} w_t \\
  =&2((w_t)^{\top}x_t^j-y_t^j)x_t^j \cdot \Delta\hat{\alpha}_{s}^i\Omega(t,s), \\
\end{aligned}
\end{equation}
where $j$ denotes the $j$-th sample for $t$-th \emph{target} node. Similarly, for the hinge loss, we can have:
 \begin{equation}\label{eq:update_x_hl}
\begin{aligned}
  &  \nabla_{(\hat{x}_s^i)}\mc{L}_t((w_t)^{\top}x_t^j,y_t^j) \\
  =&y_t^jx_t^j \cdot \nabla_{\hat{x}_s^i} w_t \\
  =&y_t^jx_t^j \cdot \Delta\alpha_{s}^i\Omega(t,s). \\
\end{aligned}
\end{equation}

Finally, the whole optimization procedure of {att}ack on {f}ederated {l}earning can be summarized in \textbf{Algorithm 1}.

\section{Experiments}\label{sec:experiment}
In this section, we evaluate the performance of the proposed poisoning attack strategies, and its convergence. More specifically, we firstly introduce several adopted datasets, followed by the experimental results. Then some analyses about our proposed model are reported.

\subsection{Real Datasets}
\begin{figure}[t]
  \centering
  \includegraphics[width=0.47\textwidth]{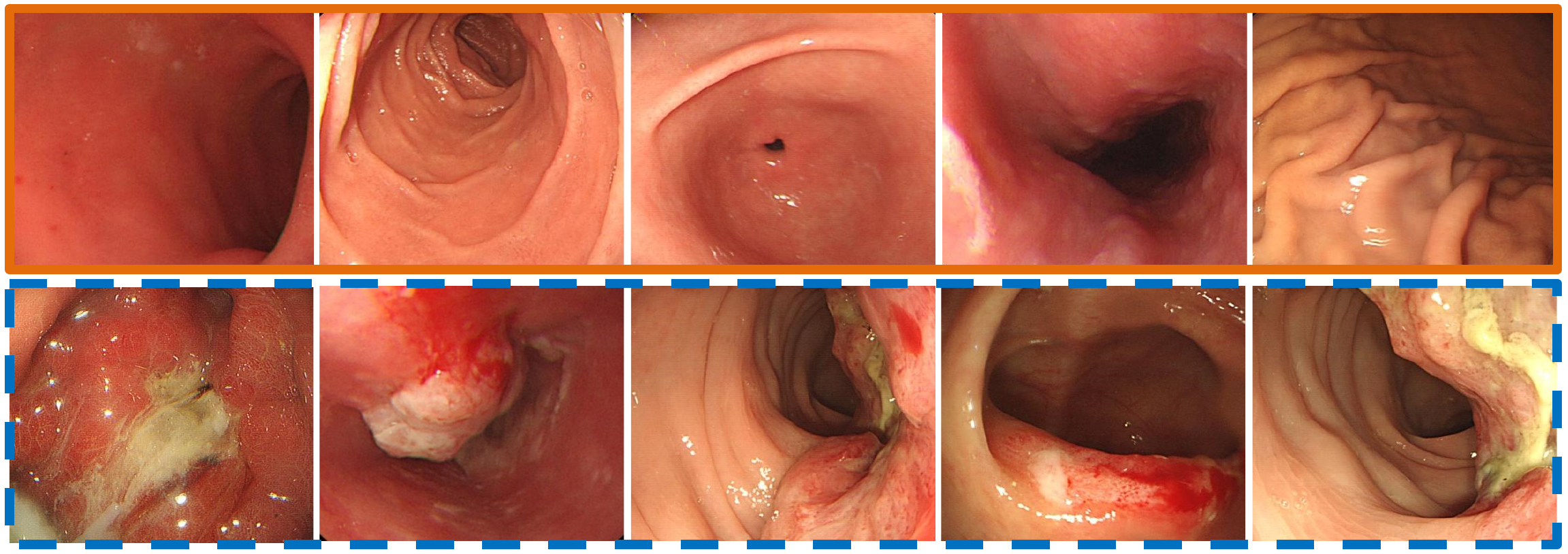}
  \caption{Sample images from \textbf{EndAD} dataset, where the first and the second rows are healthy images and lesion images, respectively.}\label{fig:EndAD_Examples}
\end{figure}

We adopt the following benchmark datasets for our experiments, containing three classification datasets and one regression dataset:

\textbf{EndAD (Endoscopic image abnormality detection):} this dataset is collected from 544 healthy volunteers and 519 volunteers with various lesions, including gastritis, cancer, bleeding and ulcer. With the help of these volunteers, we obtain 9769 lesion images and 9768 healthy images with the resolution $489\times409$. Some examples are shown in Fig.~\ref{fig:EndAD_Examples}. In the implementation, we extract a 128-dimensional deep feature via the VGG model trained in the \cite{chatfield2014return}. To simulate the federated learning procedure, we split the lesion images on a per-disease basis, while splitting the healthy images randomly. Then we obtain a federated learning dataset with 6 clinics, where each clinic can be regarded as one node, i.e., 6 nodes.

\textbf{Human Activity Recognition\footnote{https://archive.ics.uci.edu/ml/datasets/Human+Activity+ \\ Recognition+Using+Smartphones}:} this classification dataset consists of mobile phone accelerometer and gyroscope data, which are collected from 30 individuals, performing one of six activities:  \{walking,
walking-upstairs, walking-downstairs, sitting, standing, lying down\}. The provided feature vectors of time is $561$, whose variables are generated from frequency domain. In this experiment, we model each individual as a separate task and aim to predict between sitting and other activities (e.g., lying down or walking). Therefore, we have $30$ nodes in total, where each node corresponds to an individual.

\begin{table*}[t]
\caption{Comparisons between our model and state-of-the-arts in terms of classification Error and RMSE on five datasets: mean and standard errors over 10 random runs. Models with the best performance are bolded.}
\centering
\scalebox{0.80}{
\begin{tabular}{|p{70pt}|c|c|ccc|ccc|c|c|}
\hline
 {}&Metrics &Non attacks & Random direct attacks & Random indirect attacks & Random hybird attacks& Direct attacks & Indirect attacks & Hybird attacks\\
 \hline\hline
EndAD  & Error($\%$)&6.881$\pm$0.52& 7.659$\pm$1.14 & 6.888$\pm$0.45 & 7.154$\pm$0.16 &\textbf{28.588$\pm$3.74}      &7.324$\pm$0.62      & 16.190$\pm$2.26 \\ \hline
Human Activity  & Error($\%$)&2.586$\pm$0.84 & 3.275$\pm$0.71  & 2.894$\pm$0.83 & 3.172$\pm$0.69 & \textbf{29.422$\pm$2.96} &3.438$\pm$0.34 & 17.829$\pm$2.75  \\ \hline
Landmine  & Error($\%$)&5.682$\pm$0.28 &5.975$\pm$0.36    &5.735$\pm$0.36  &5.819$\pm$0.22  & \textbf{13.648$\pm$0.54} &7.428$\pm$0.39  & 9.579$\pm$0.27  \\ \hline

&Avg. Error ($\%$) & 5.049$\pm$0.55  & 5.636$\pm$0.74 & 5.172$\pm$0.55  & 5.382$\pm$0.36 &   \textbf{23.886$\pm$2.41}   & 6.069$\pm$0.45  & 14.533$\pm$1.76   \\  \hline\hline

Parkinson-Total    &RMSE& 6.302$\pm$0.45 & 13.651$\pm$2.10 & 6.633$\pm$0.75 &11.145$\pm$1.83            & \textbf{44.939$\pm$3.21}      &7.763$\pm$0.82 &21.990$\pm$3.17 \\ \hline
Parkinson-Moter    &RMSE&  4.125$\pm$0.50 & 11.472$\pm$2.51 & 5.046$\pm$1.14 & 9.422$\pm$1.81 & \textbf{32.992$\pm$3.78}            &6.866$\pm$1.21      & 16.956$\pm$3.78 \\ \hline
&Avg. RMSE($\%$) &  5.213$\pm$0.48 & 12.562$\pm$2.31 & 5.839$\pm$0.95  & 10.284$\pm$1.82 & \textbf{38.966$\pm$3.49}   & 7.314$\pm$1.02  &19.473$\pm$3.48   \\  \hline
 \end{tabular}
}
\label{table:result}
\end{table*}

\textbf{Landmine:} this dataset intends to detect whether or not a land mine is presented in an area based on radar images. Moreover, it is modeled as a binary classification problem. Each object in a given dataset is represented by a 9-dimensional feature vector (four-moment based features, three correlation-based features, one energy-ratio feature, one  spatial variance feature) and the corresponding binary label (1 for landmine and -1 for clutter). This dataset consists of a total of 14,820 samples divided into 29 different geographical regions, i.e., the total node number is 29.


\textbf{Parkinson Data}\footnote{https://archive.ics.uci.edu/ml/datasets/parkinsons+telemonitoring} is used to predict Parkinson’s disease symptom score for patients based on 16 biomedical features. The Parkinson dataset contains 5,875 observations for 42 patients, and predicting the symptom score for each patient is treated as a regression task, leading to 42 regression tasks with the number of instances for each task ranging from 101 to 168. Additionally, the dataset's output is a score consisting of Total and Motor, we thus have two regression datasets: \textbf{Parkinson-Total} and \textbf{Parkinson-Motor} in this experiment.

For the evaluation, we adopt the classification Error to evaluate the classification performance. For the regression problems, RMSE (root mean squared error) is a reasonable evaluation metric. The smaller the Error and RMSE value is, the better the performance of model will be, \emph{i.e.,} the weaker the attack strategy will be.

\subsection{Evaluating Attack Strategies}
In our experiments, we evaluate the performance of our proposed poisoning attack strategies with random/direct/inditect/hybrid attacks on different datasets. For each dataset, we randomly choose half of nodes as $\mc{N}_{\mr{tar}}$, while selecting the rest of nodes as free nodes. For example, the number of $\mc{N}_{\mr{tar}}$ for Human Activity dataset is 15, and the number of $\mc{N}_{\mr{tar}}$ for EndAD dataset is 3, respectively. Moreover, the details of evaluated attack strategies are:
\begin{itemize}
\setlength{\itemsep}{1pt}
\setlength{\parsep}{0pt}
\setlength{\parskip}{0pt}
  \item Direct attacks: all the source attacking nodes $\mc{N}_{\mr{sou}}$ are set as the same as the target nodes, \emph{i.e.,} $\mc{N}_{\mr{sou}}=\mc{N}_{\mr{tar}}$.
  \item Indirect attacks: all the source attacking nodes $\mc{N}_{\mr{sou}}$ are from the rest of nodes, where the number of $\mc{N}_{\mr{sou}}$ is same as that of $\mc{N}_{\mr{tar}}$.
  \item Hybrid attacks: all the source attacking nodes $\mc{N}_{\mr{sou}}$ are randomly selected from all the nodes, where the number of $\mc{N}_{\mr{sou}}$ is same as that of $\mc{N}_{\mr{tar}}$.
  \item Random  Direct/Indirect/Hybrid attacks: the attack strategy are same as that of Direct/Indirect/Hybrid attacks. However, the injected data samples are randomly chosen.
\end{itemize}

\begin{figure}[t]
\center
    \subfigure[EndAD]{
        \begin{minipage}[a]{0.24\textwidth}
          \centering
    \includegraphics[width =112pt ,height =72pt]{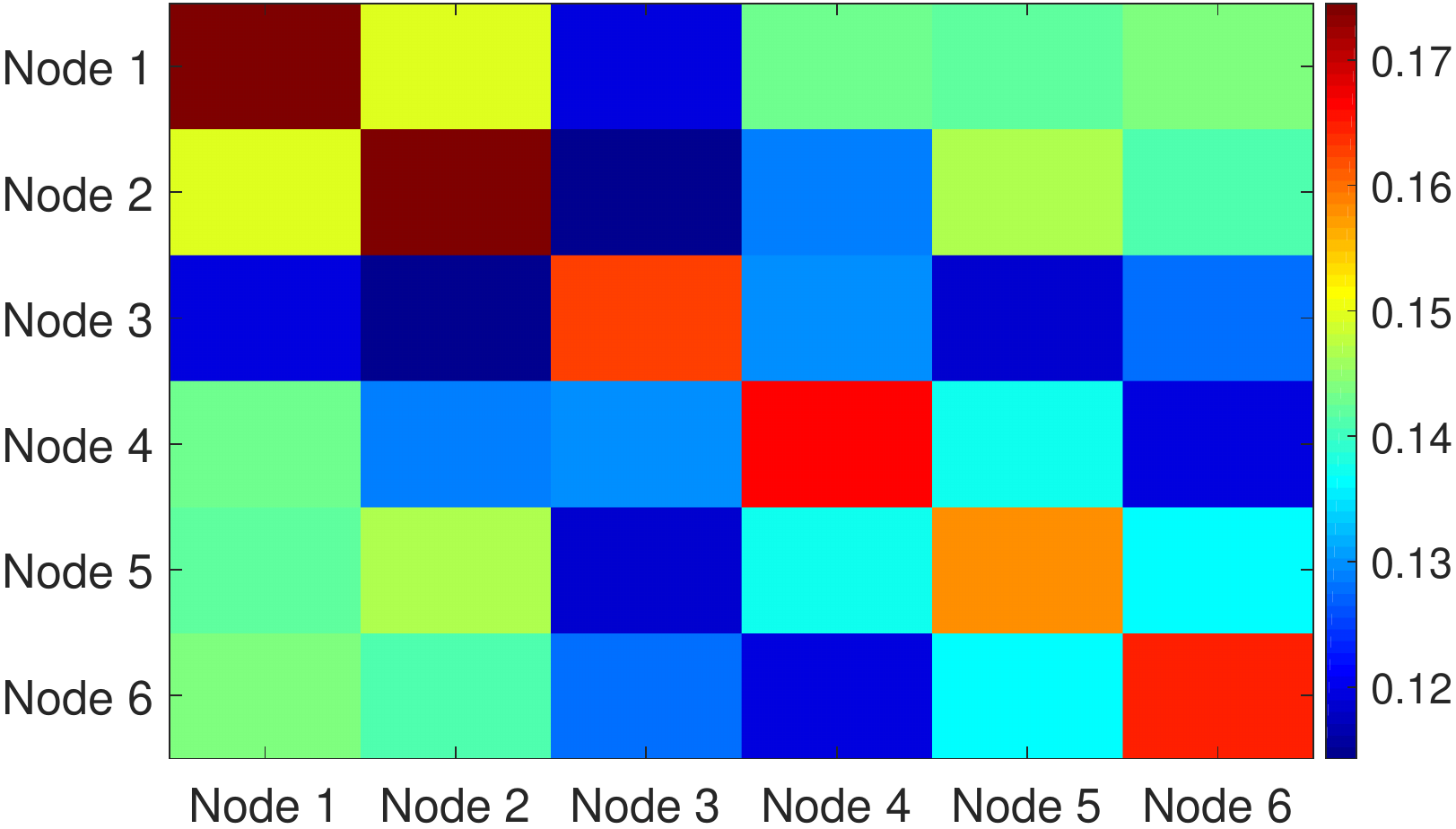}
         \end{minipage}}
         \hspace{-5.0mm}
     \subfigure[Human Activity]{
        \begin{minipage}[a]{0.24\textwidth}
          \centering
    \includegraphics[width =112pt ,height =72pt]{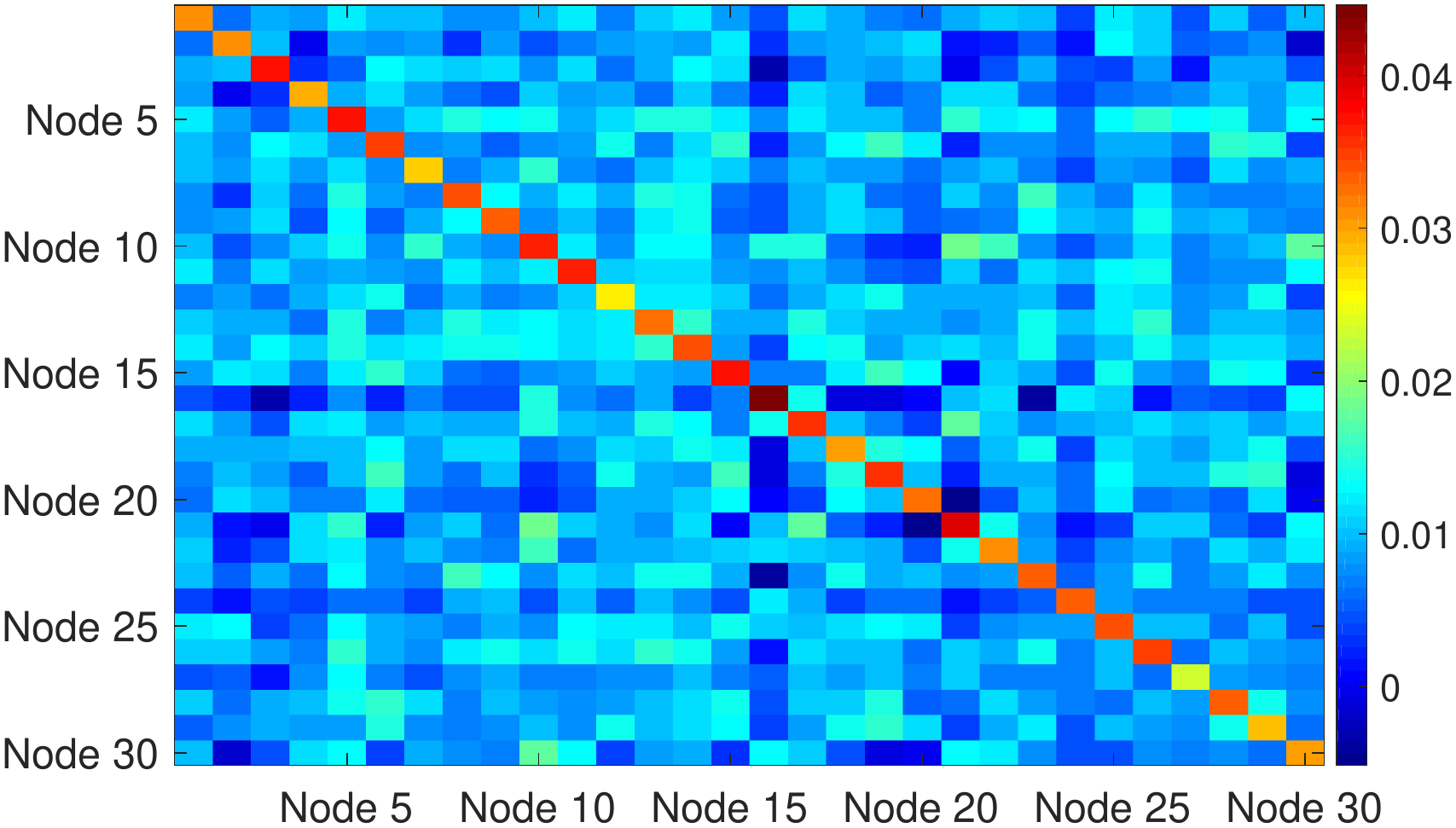}
        \end{minipage}}
        \hspace{-2.4mm} \\
        \subfigure[Landmine]{
        \begin{minipage}[a]{0.24\textwidth}
          \centering
    \includegraphics[width =112pt ,height =72pt]{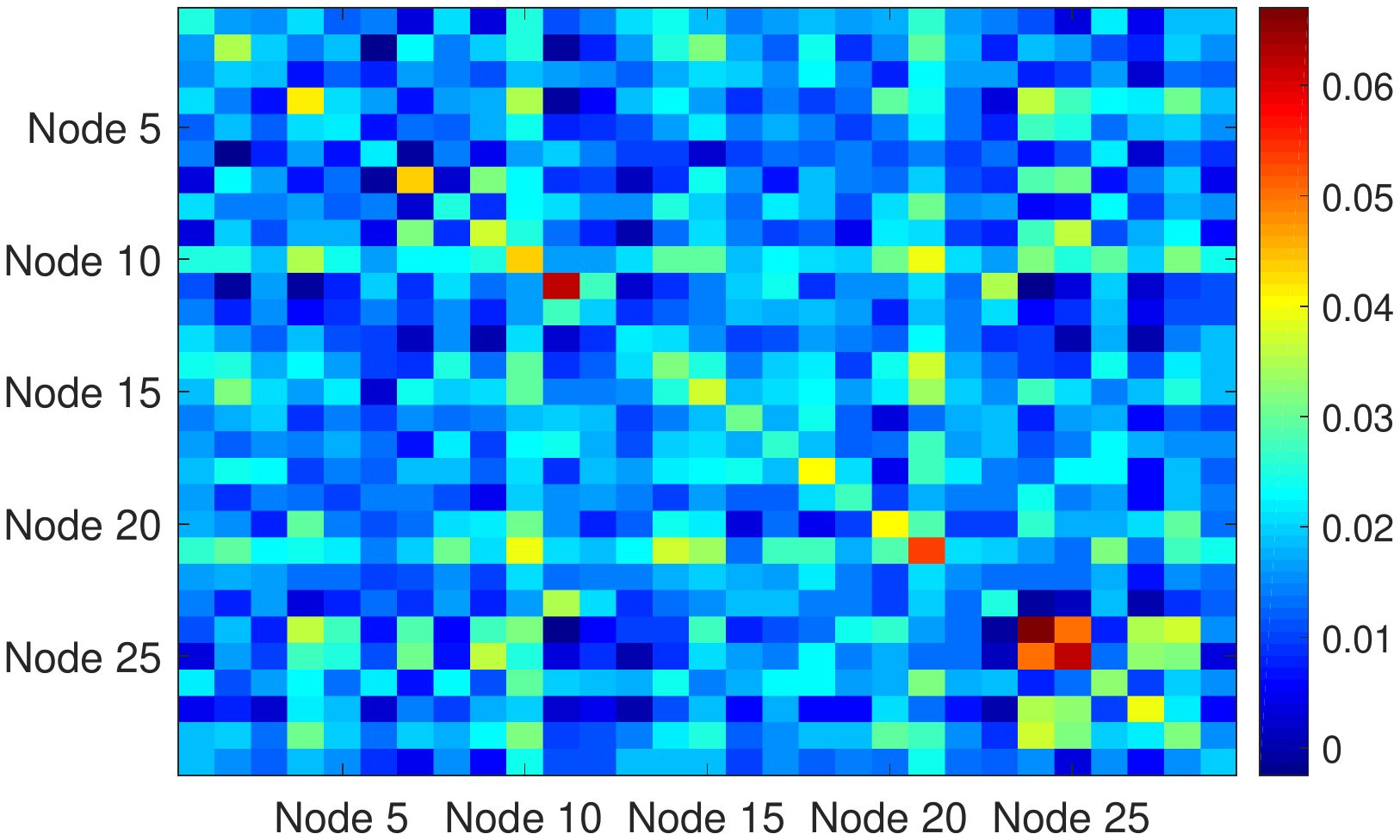}
        \end{minipage}}
        \hspace{-4.4mm}
    \subfigure[Parkinson]{
        \begin{minipage}[a]{0.24\textwidth}
          \centering
    \includegraphics[width =112pt ,height =72pt]{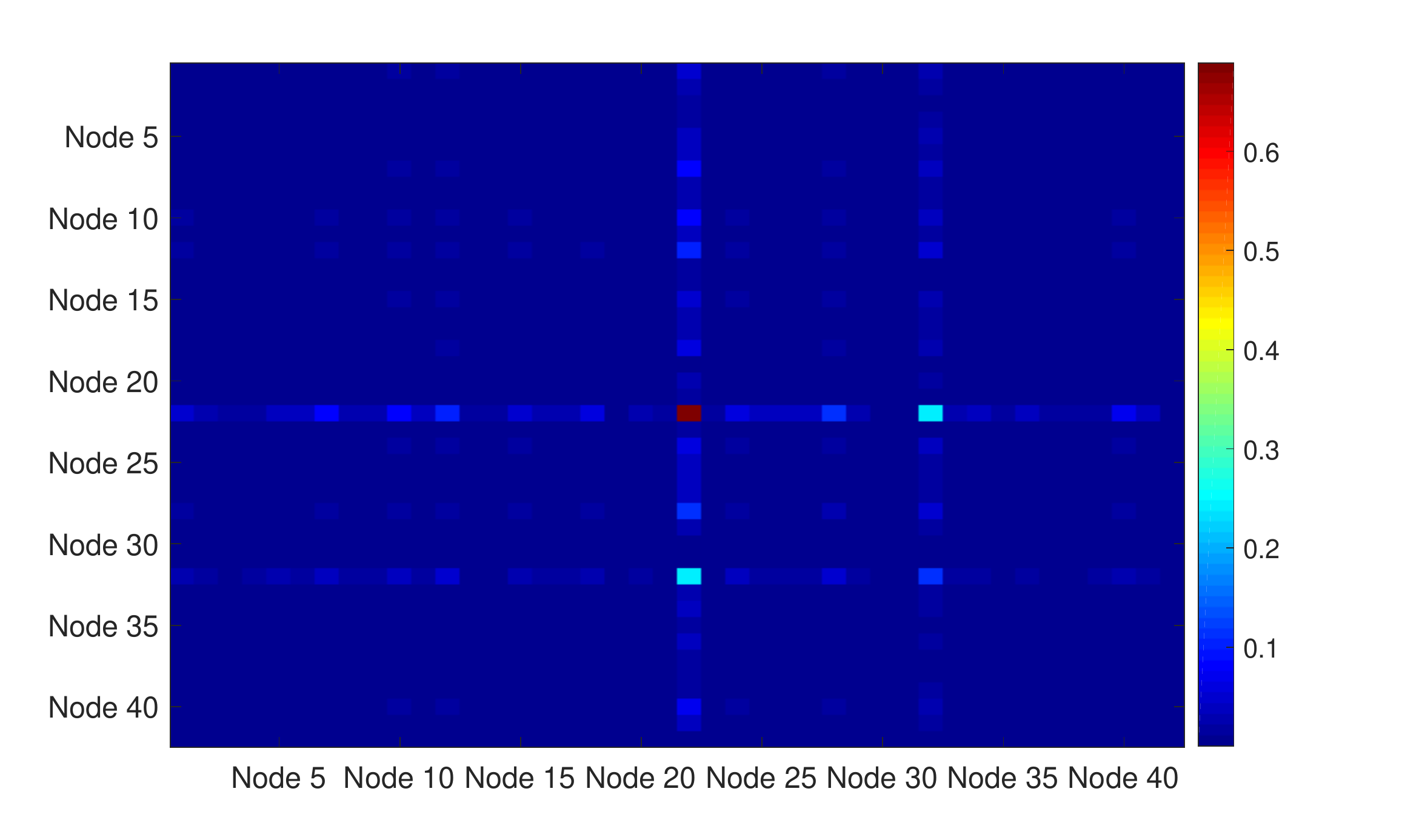}
        \end{minipage}}
        \hspace{-2.4mm}
   \caption{The correlation matrices $\Omega$ of different datasets: End$\_$AD dataset (a), Human Activity dataset (b), Landmine dataset (c) and Parkinson dataset (d), where the first half of nodes and the rest of nodes denote the $\mc{N}_{\mr{tar}}$ and $\mc{N}_{\mr{sou}}$ for the indirect attack scenario, respectively. The darker color indicates the higher correlation for each dataset, and vice versa.}
  \label{fig:CorrelationMatrix}
\end{figure}

For the used parameters in our model, the step size $\eta_1$ in Eq.~\eqref{eq:update_x} is set as $100$, both $\lambda_1$ and $\lambda_2$ are set as $0.001$ among all the experiments for a fair comparison. Furthermore, the number of injected data samples for each source attacking node is set as $20\%$ of the clean data. The experimental results averaged over ten random repetitions are presented in Table.~\ref{table:result}. From the presented results, we have the following observations:
\begin{itemize}
\setlength{\itemsep}{2.5pt}
\setlength{\parsep}{0pt}
\setlength{\parskip}{0pt}
  \item All the attack strategies (\emph{e.g.,} direct attacks, random attacks, etc) have the impact on all the datasets, which verifies the vulnerability of federated machine learning. Among all the attack strategies, notice that the direct attacks can significantly damage the classification or regression on all the datasets. For example, the performances of EndAD and Human Activity datasets can lead to $21.707\%$ and $26.836\%$ deterioration in terms of classification Error, respectively. This observation indicates direct attacks are the big threats to the federated machine learning system.
\item When comparing with the random attack strategies, our proposed attack strategies can obtain better deterioration performance among most cases for the federated machine learning problems, which justifies that the learning attack strategies can work better than just launching random attack strategies.
 \item For the indirect attack strategy, we can notice its performances are not good as direct and hybrid attack strategies, \emph{e.g.,} Human Activity Recognition dataset. This is because that the indirect attack can be successfully launched via effectively using the communication protocol among different nodes, where the communication protocol is bridged by the model relationship matrix $\Omega$ in this paper. To verify this observation, we also present the corresponding correlation matrix of each used dataset in Figure~\ref{fig:CorrelationMatrix}. As illustrated in Figure~\ref{fig:CorrelationMatrix}, the nodes from $\mc{N}_{\mr{tar}}$ and $\mc{N}_{\mr{sou}}$ have highly correlated in Landmine dataset, \emph{i.e.,} Figure~\ref{fig:CorrelationMatrix}(c). However, the node correlations in the Parkinson dataset (\emph{i.e.,} Figure~\ref{fig:CorrelationMatrix}(d)) are not close. This observation indicates the reason why indirect attack strategy performs better on the Landmine dataset when comparing with other used datasets.
\end{itemize}

\subsection{Sensitivity Study and Convergence Analysis}
This subsection first conducts sensitivity studies on how the ratio of injected data and step size $\eta$ affect different attack strategies. Then we also provide the convergence analysis about our proposed AT$^2$FL.

\subsubsection{Effect of Ratio of Injected Data}
\begin{figure}[t]
\center
\hspace{-5pt}
    \subfigure[EndAD]{
        \begin{minipage}[a]{0.24\textwidth}
          \centering
    \includegraphics[width =114pt ,height =82pt]{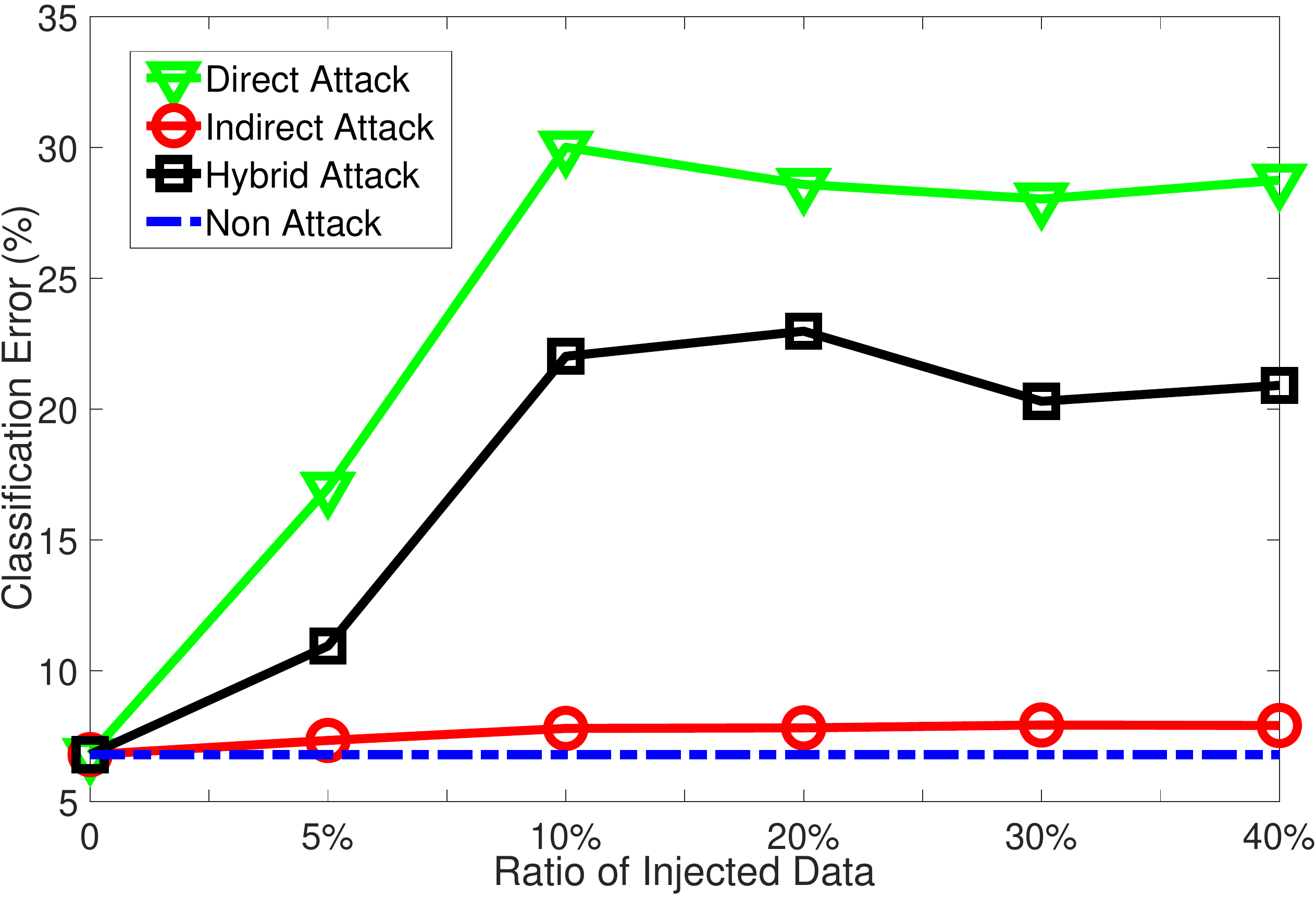}
         \end{minipage}}
         \hspace{-4.4mm}
     \subfigure[Human Activity]{
        \begin{minipage}[a]{0.24\textwidth}
          \centering
    \includegraphics[width =114pt ,height =82pt]{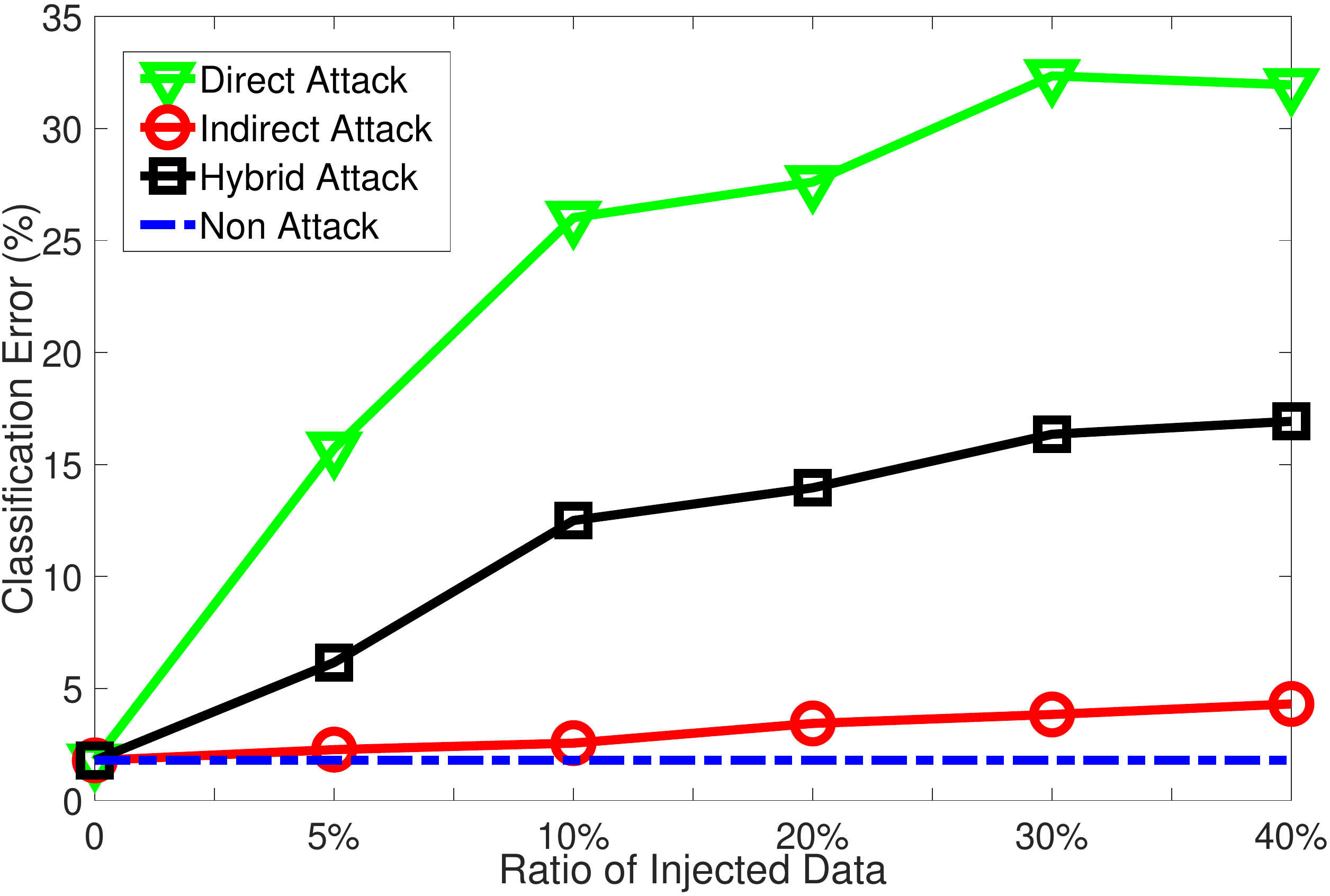}
        \end{minipage}}
        \hspace{-5.4mm} \\
        \subfigure[Landmine]{
        \begin{minipage}[a]{0.24\textwidth}
          \centering
           \hspace{-6pt}
    \includegraphics[width =114pt ,height =82pt]{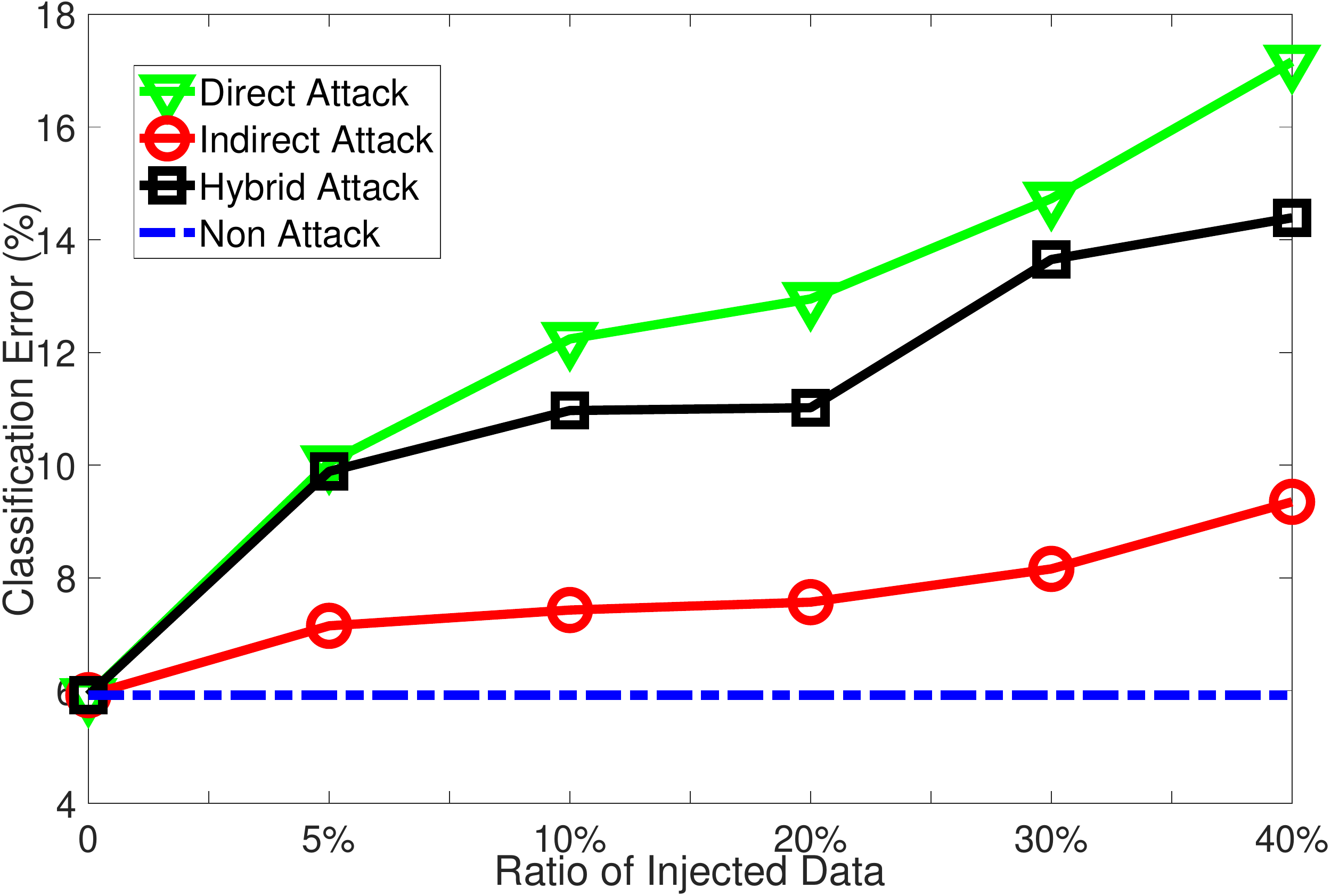}
        \end{minipage}}
        \hspace{-5.4mm}
    \subfigure[Parkinson-Total]{
        \begin{minipage}[a]{0.24\textwidth}
          \centering
    \includegraphics[width =114pt ,height =82pt]{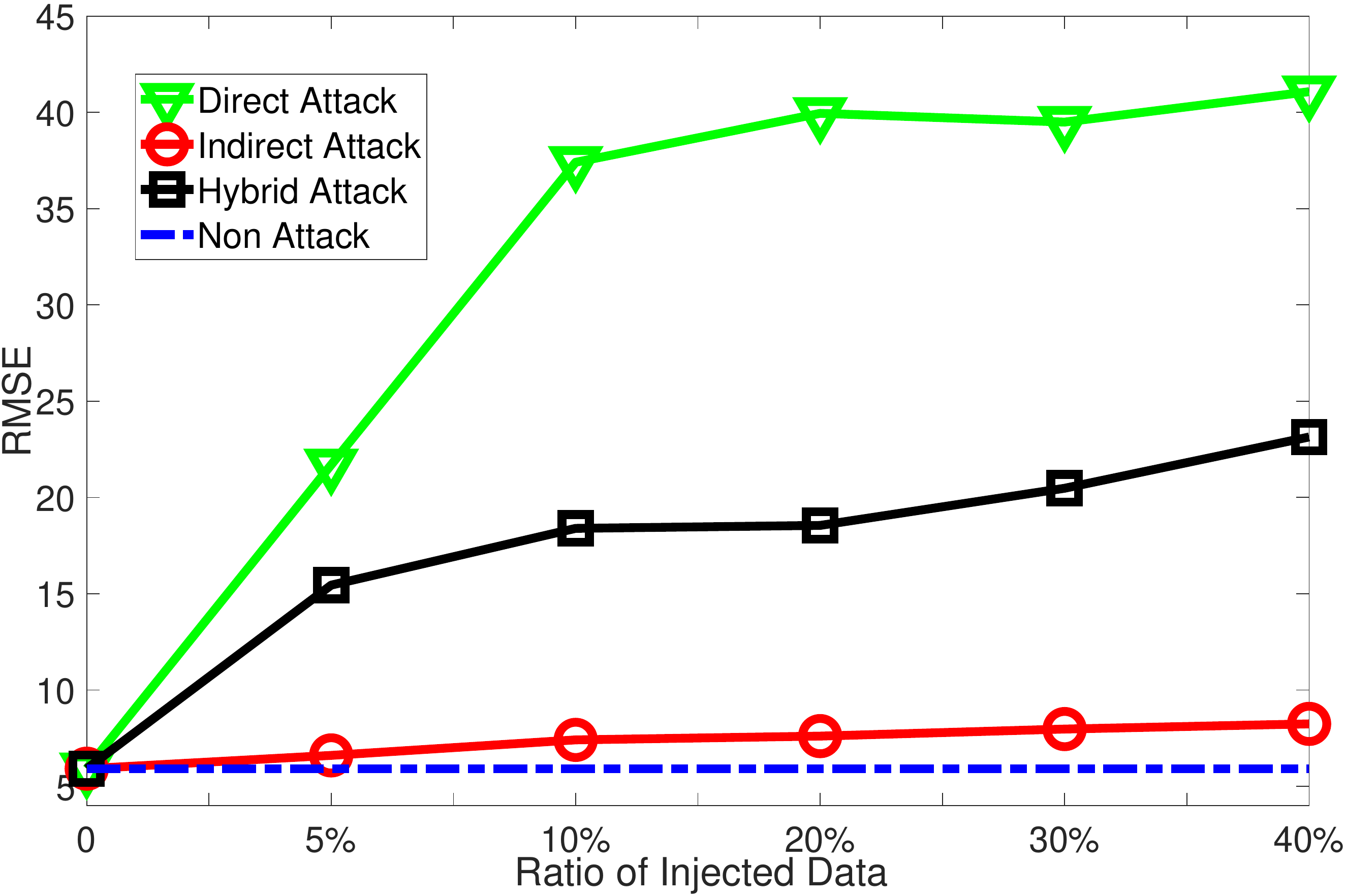}
        \end{minipage}}
        \hspace{-2.4mm}
   \caption{The effect of ratio of injected data on EndAD (a), Human Activity (b), Landmine (c) and Parkinson-Total (d) datasets, where different lines denote different types of attacking strategies.}
  \label{fig:inject}
\end{figure}

To investigate the effect of ratio of injected poisoned data, we use all the datasets in this subsection. For each dataset, we take the same attack setting as that in Table~\ref{table:result}, \emph{e.g.,} for the direct attack, the $\mc{N}_{\mr{tar}}$ and $\mc{N}_{\mr{sou}}$ are selected by randomly selecting half of nodes as $\mc{N}_{\mr{tar}}$, and the rest of nodes to form $\mc{N}_{\mr{sou}}$. For the injected data in each source attacking node, we tune it in a range $\{0,5\%,10\%,20\%,30\%,40\%\}$ of the clean data, and present the results in Figure~\ref{fig:inject}. From the provided results in Figure~\ref{fig:inject}, we can find that: 1) For all the used datasets, the performances under different attack strategies are decreased with the increasing of the injected data. This observation demonstrates the effectiveness of attack strategies computed by our proposed AT$^2$FL; 2) Obviously, the performances of direct attack are better than that of indirect attack by varying the ratio of injected data. This because that it can directly involve the data of $\mc{N}_{\mr{tar}}$; 3) Indirect attack almost has no effect on EndAD dataset since the values of correlation matrix $\Omega$ for this dataset are relatively low, and each node can learn good classifier without the help of other nodes; 4) Different from other datasets, the classification Errors slightly decrease after the ratio of injected data is $10\%$. It is because the classification Error is computed on the test data while the formulation in Eq~\eqref{eq:ours} aims to maximize the loss on the training data. Meanwhile, the classification Error on the test data obtains the upper bound when the ratio of injected data is $10\%$.

\begin{figure}[t]
\center
\hspace{-5pt}
    \subfigure[Human Activity]{
        \begin{minipage}[a]{0.24\textwidth}
          \centering
    \includegraphics[width =114pt ,height =82pt]{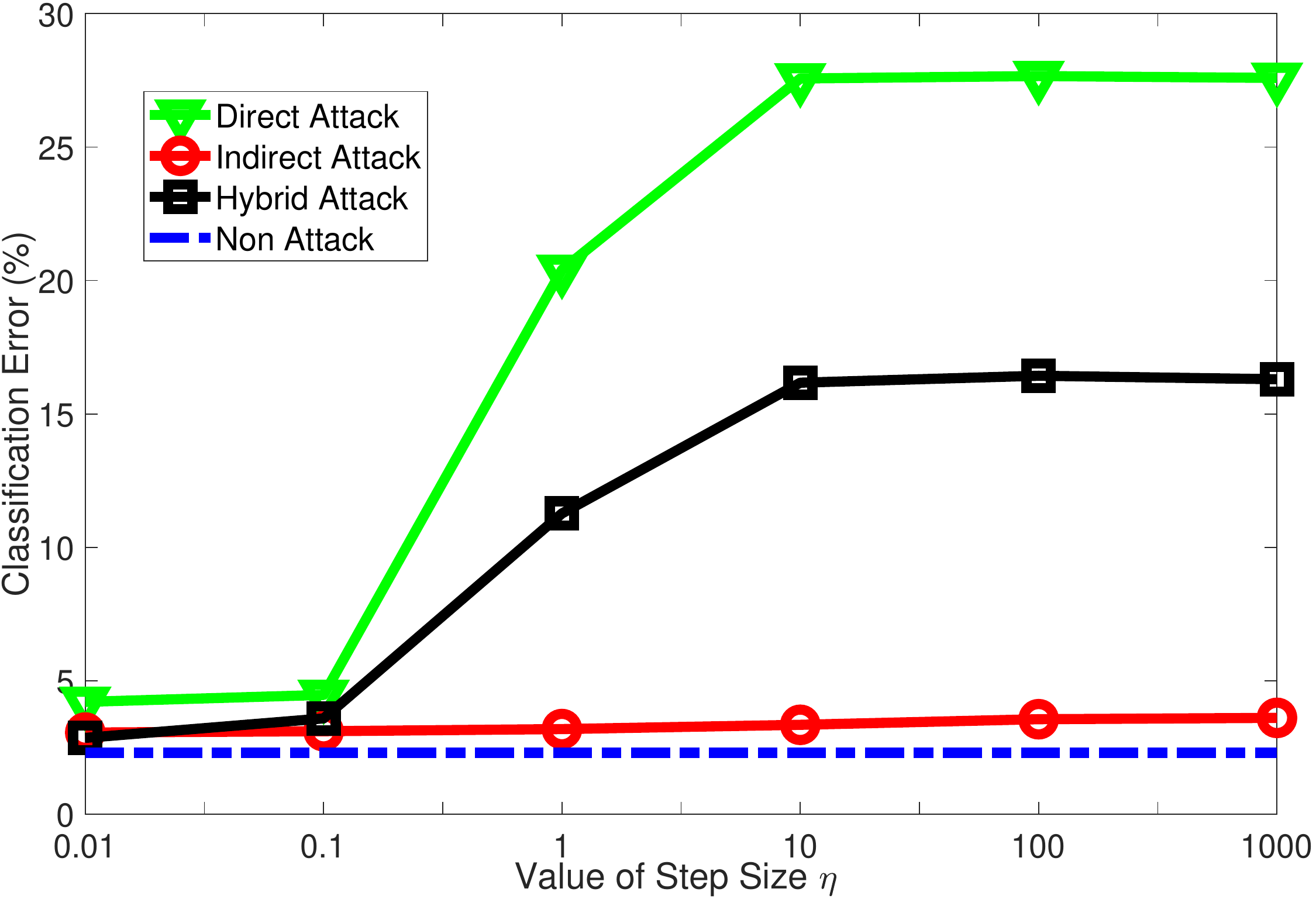}
         \end{minipage}}
         \hspace{-4.4mm}
     \subfigure[Landmine]{
        \begin{minipage}[a]{0.24\textwidth}
          \centering
    \includegraphics[width =114pt ,height =82pt]{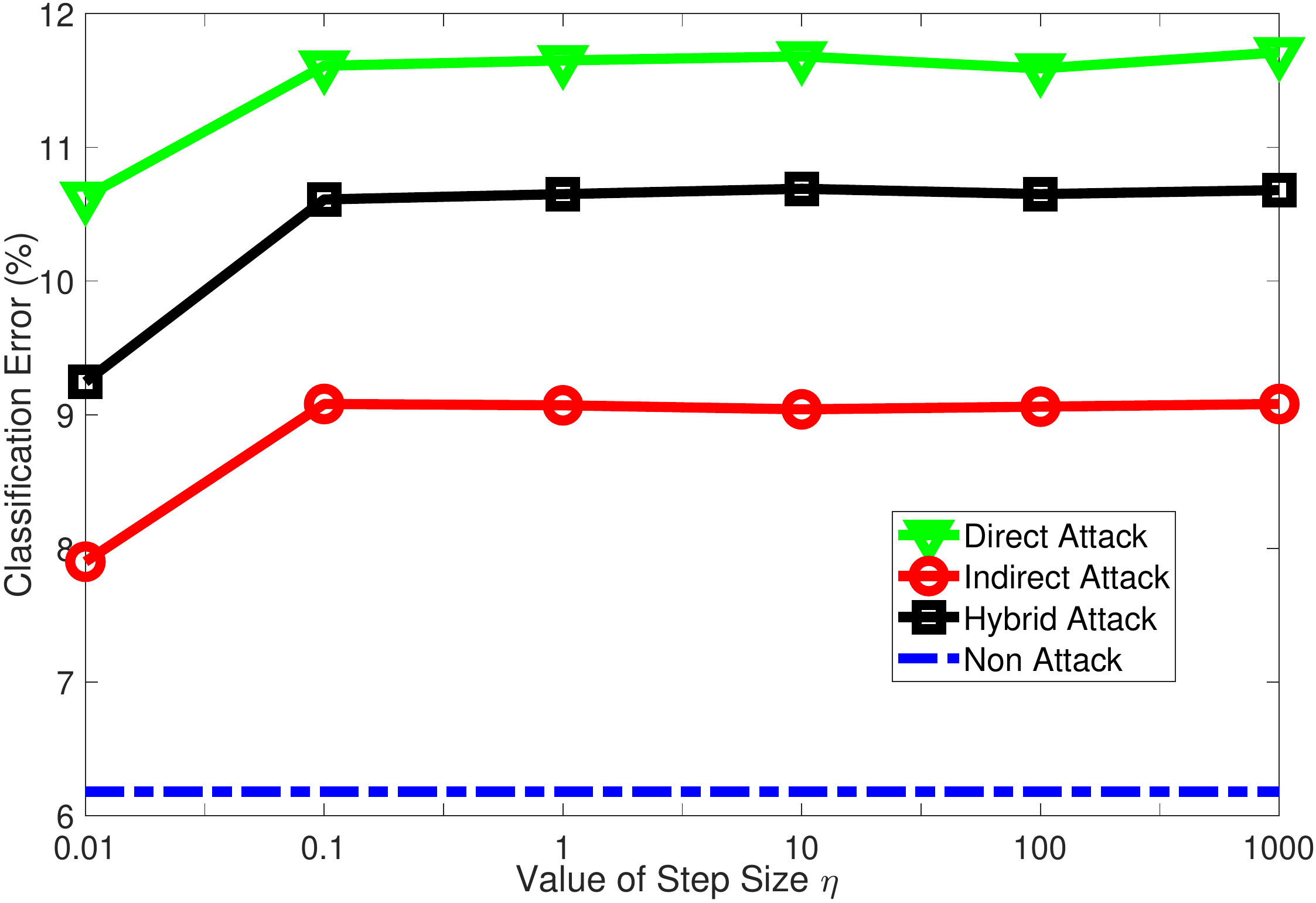}
        \end{minipage}}
        \hspace{-3.4mm}
   \caption{The effect of step size $\eta$ of Eq.~\eqref{eq:update_x} on Human Activity (a) and Landmine (b) datasets, where different lines denote different attacking strategies.}
  \label{fig:stepsize}
\end{figure}

\subsubsection{Effect of Step Size $\eta$}
In order to study how the value of step size $\eta$ affects the performance of our proposed attack strategies, we adopt Human Activity Recognition and Landmine datasets in this subsection. By fixing other parameters (\emph{e.g.,} $\lambda_1$ and $\lambda_2$) and varying the value of step size $\eta$ of Eq.~\eqref{eq:update_x} in range $\{0.01,0.1,1,10,100,1000\}$, we present the corresponding 20-th iteration performances of different attack strategies in Figure~\ref{fig:stepsize}. Notice that the performances of different attack strategies are improved with the increasing of the step size $\eta$. This is because the gradient in Eq.~\eqref{eq:update_x} with small step size will be treated as noise information, which will just have little impact on the classification performance. However, these attack strategies also outperforms the non-attack strategy. Furthermore, the performance of different attack strategies tend to a fixed point with the increasing of step size $\eta$. This observation indicates that our AT$^2$FL can effectively converge to a local optimum when the step size is enough.

\begin{figure}[t]
\center
    \subfigure[EndAD]{
        \begin{minipage}[a]{0.24\textwidth}
          \centering
    \includegraphics[width =110pt ,height =85pt]{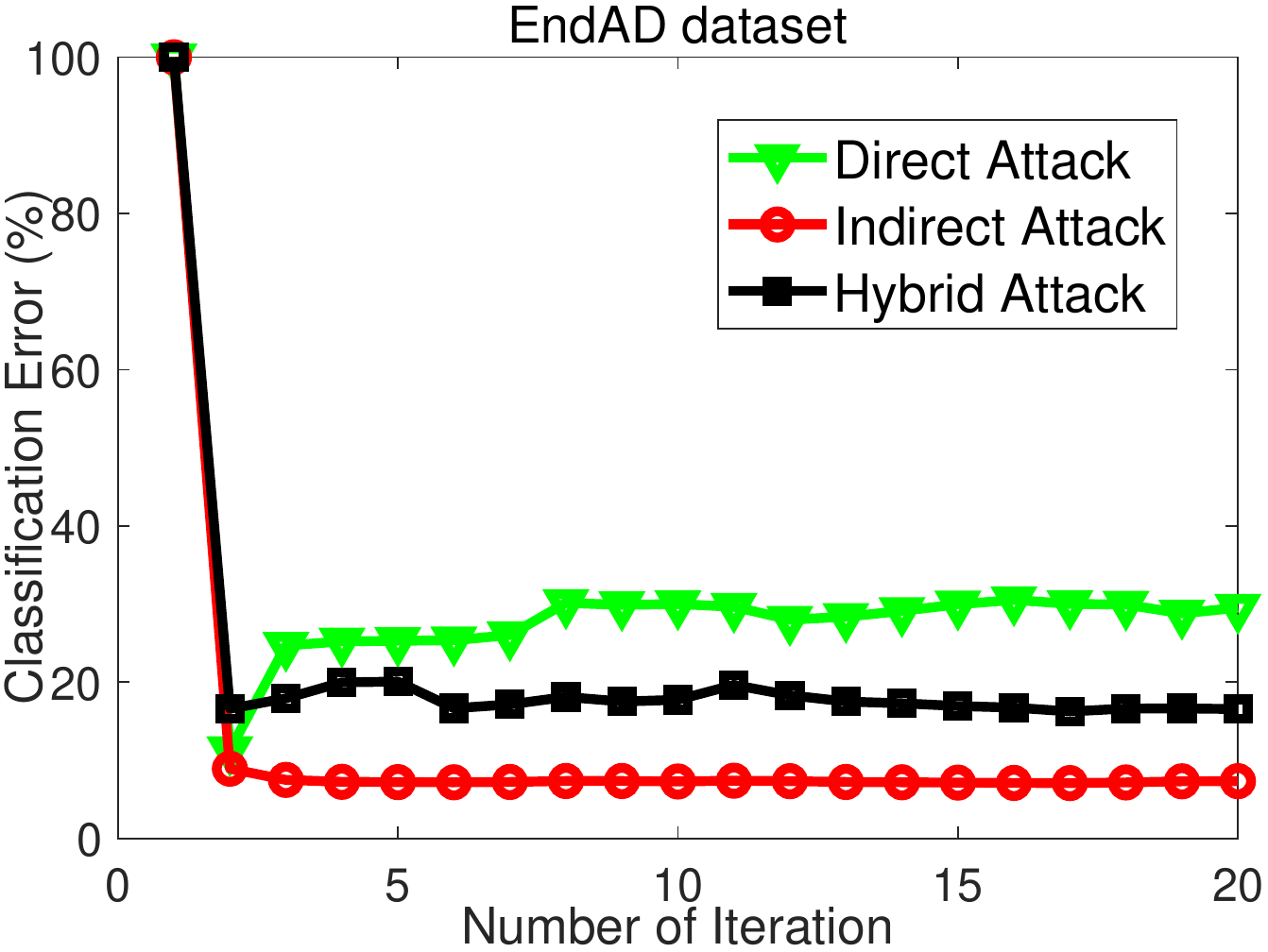}
         \end{minipage}}
         \hspace{-5.4mm}
     \subfigure[Human Activity]{
        \begin{minipage}[a]{0.24\textwidth}
          \centering
    \includegraphics[width =114pt ,height =85pt]{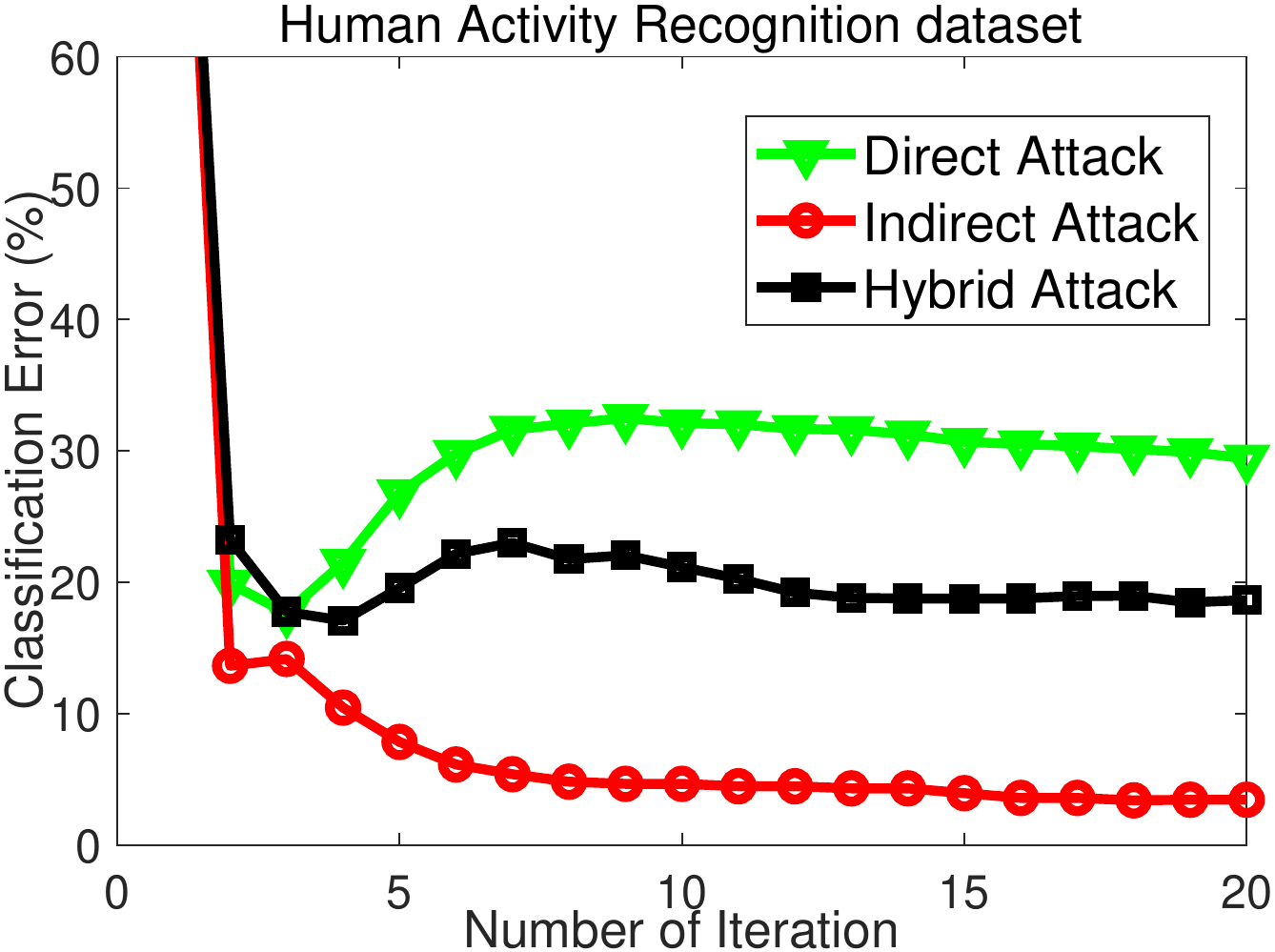}
        \end{minipage}}
        \hspace{-3.4mm} \\
        \subfigure[EndAD]{
        \begin{minipage}[a]{0.24\textwidth}
          \centering
    \includegraphics[width =110pt ,height =85pt]{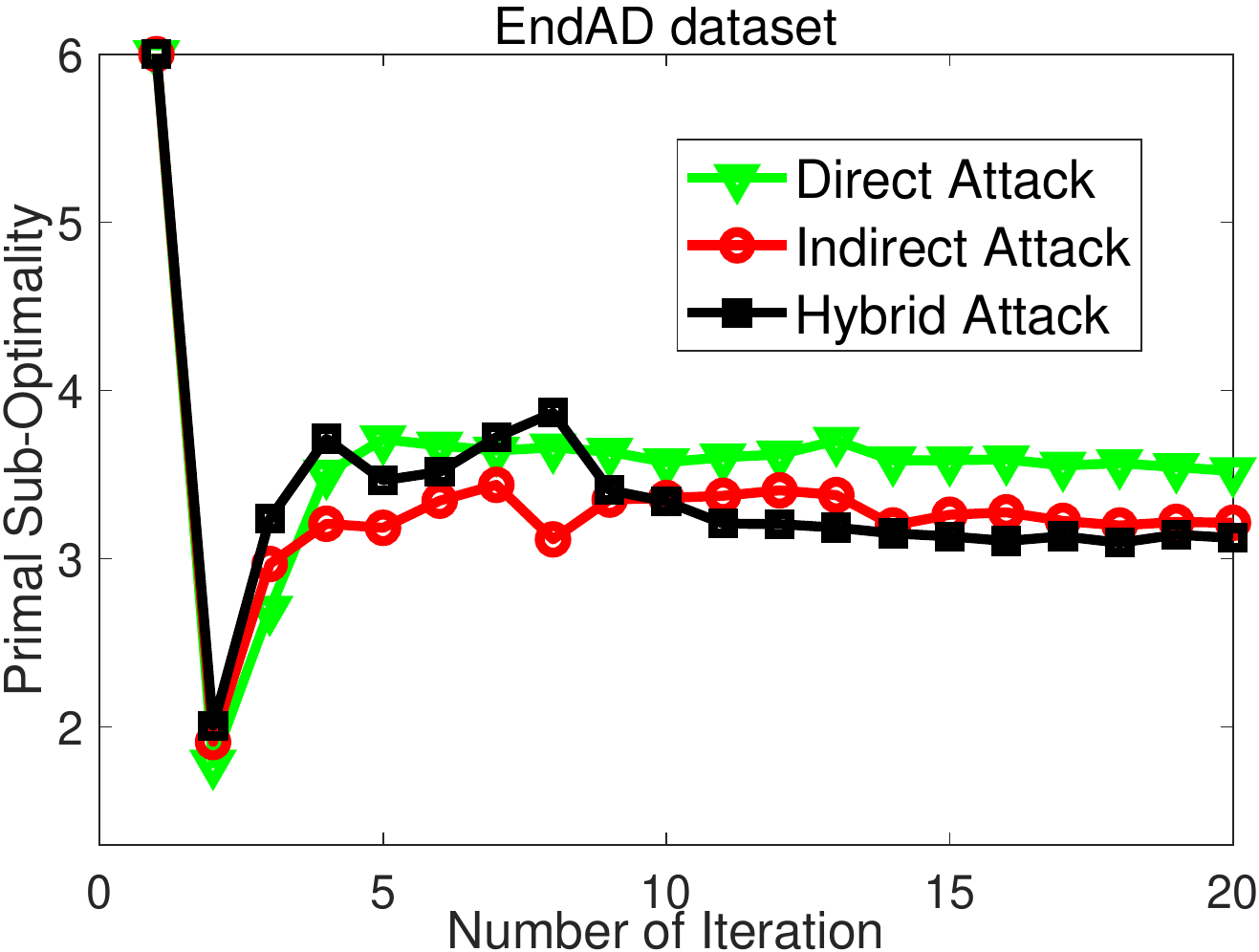}
        \end{minipage}}
        \hspace{-5.4mm}
    \subfigure[Human Activity]{
        \begin{minipage}[a]{0.24\textwidth}
          \centering
    \includegraphics[width =110pt ,height =85pt]{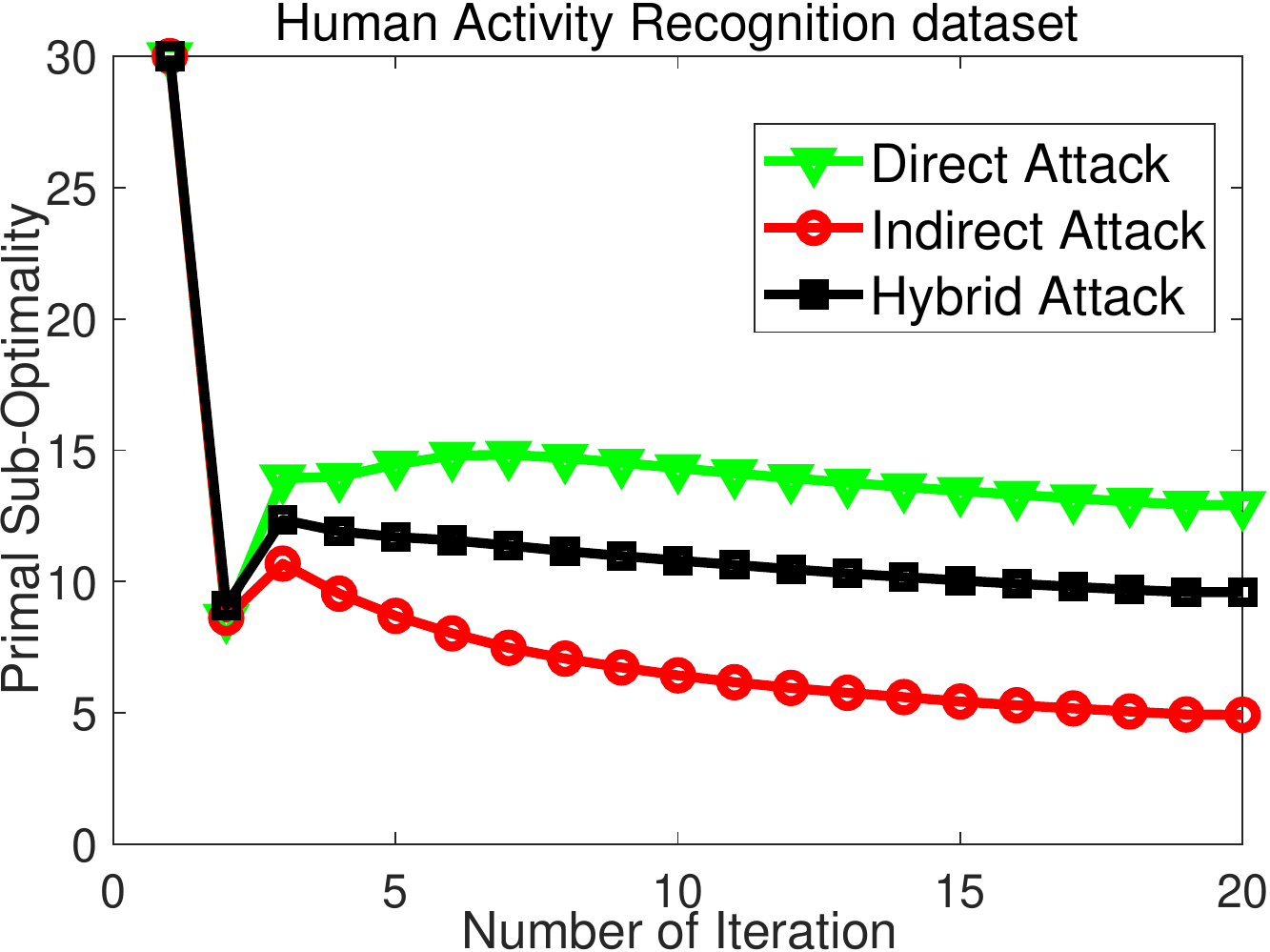}
        \end{minipage}}
        \hspace{-2.4mm}
   \caption{The convergence curves of our proposed AT$^2$FL algorithm, where the classification Errors are in (a) EndAD and (b) Human Activity datasets, and the primal sub-optimality performances of the lower level problem for Eq.~\eqref{eq:ours} are in (c) EndAD and (d) Human Activity datasets.}
  \label{fig:convergence}
\end{figure}

\subsubsection{Convergence of Proposed AT$^2$FL}
To study the convergence of our proposed AT$^2$FL algorithm, we adopt the EndAD and Human Activity Recognition datasets in this subsection. Specifically, under different attack strategies, we present the primal sub-optimality of the lower level problem for Eq.~\eqref{eq:ours} and classification Error ($\%$) of the nodes in $\mc{N}_{\mr{tar}}$ in Figure~\ref{fig:convergence}. From the presented curves in Figure~\ref{fig:convergence}, we can notice that the original values of primal sub-optimality and classification error are higher. This is because that the weight matrix $W$ in Eq.~\eqref{eq:ours} is just initialized with the injected data points. Then our proposed AT$^2$FL algorithm can converge to a local optima after a few iterations for all the three kinds of attacks on EndAD and Human Activity Recognition datasets. This observation indicates the effectiveness of our proposed AT$^2$FL algorithm.

\section{Conclusion} \label{sec:conclusion}
In this paper, we take an earlier attempt on how to effectively launch data poisoning attacks on federated machine learning. Benefitting from the communication protocol, we propose a bilevel data poisoning attacks formulation by following general data poisoning attacks framework, where it can include three different kinds of attacks. As a key contribution of this work, we design a \underline{ATT}ack on \underline{F}ederated \underline{L}earning (AT$^2$FL) to address the system challenges (\emph{e.g.,} high communication cost) existing in federated setting, and further compute optimal attack strategies. Extensive experiments demonstrate that the attack strategies computed by AT$^2$FL can significantly damage performances of real-world applications. From the study in this paper, we find that the communication protocol in federated learning can be used to effectively launch indirect attacks, \emph{e.g.,} when two nodes have strong correlation. Except for the horizontal federated learning in this work, we will consider the data poisoning attacks study on vertical (feature-based) federated learning and federated transfer learning in the future.


\appendices
\section{Definition of Least-square and Hinge Losses}
 For the regression problem in this paper, we adopt least-square loss: $\mc{L}_{\ell}(w_{\ell}^{\top}x_{\ell}^i)=(w_{\ell}^{\top}x_{\ell}^i-y_{\ell}^i)^2$, and dual formulation is:
     \begin{equation}\label{eq:dual_squaredloss}
       \mc{L}_{\ell}^*(-\alpha_{\ell}^i)=-y_{\ell}^i\alpha_{\ell}^i+(\alpha_{\ell}^i)^2/4,
     \end{equation}
   For the classification problems, we adopt hinge loss: $\mc{L}_{\ell}(w_{\ell}^{\top}x_{\ell}^i)=\mr{max}(0,1-w_{\ell}^{\top}x_{\ell}^iy_{\ell}^i)$, and the dual problem is:
      \begin{equation}\label{eq:dual_hingeloss}
       \mc{L}_{\ell}^*(-\alpha_{\ell})=\begin{cases}
       -y_{\ell}^i\alpha_{\ell}^i,  \quad\quad  0\leq y_{\ell}^i\alpha_{\ell}^i\leq 1, \\
        \infty  \quad\quad\quad\quad\quad  \mr{otherwise},
       \end{cases}
     \end{equation}
where $\alpha_{\ell}^i$ is the corresponding dual variable for the $\ell$-th node in the regression or classification problems.




\ifCLASSOPTIONcaptionsoff
  \newpage
\fi



%

\bibliographystyle{plain}
\bibliography{Multitask}

\end{document}